%% file: main.tex
\documentclass[a4paper,12pt]{article}
\pdfoutput=1
\usepackage{jcappub}             
\usepackage{amsmath,amssymb}
\usepackage{graphicx}
\usepackage[utf8]{inputenc}
\usepackage{cancel}
\usepackage[colorlinks=true]{hyperref}
\usepackage{color}
\usepackage{array,multirow}
\usepackage{subfigure}
\usepackage{float}
\usepackage{comment}
\usepackage{tikz}
\usepackage{booktabs}
\usetikzlibrary{positioning}
\usetikzlibrary{decorations.pathmorphing}

\title{Dark Matter as a Source for Lepton Flavor Violation}

\author[a]{Jeremy Echeverria,}
\author[b]{Patricio Escalona,}
\author[c,d,e,f]{Farinaldo Queiroz,}
\author[c]{David Suarez}

\affiliation[a]{Departamento de Ciencias Fisicas, Universidad Andres Bello, 
Sazi\'e 2212, Piso 7, Santiago, Chile}

\affiliation[b]{Departamento de Física, Universidade Federal da Paraíba, 58051-970, João Pessoa, PB, Brazil}

\affiliation[c]{International Institute of Physics, Universidade Federal do Rio Grande do Norte, Campus Universit\'ario, Lagoa Nova, Natal-RN 59078-970, Brazil}

\affiliation[d]{Departamento de F\'isica, Universidade Federal do Rio Grande do Norte, 59078-970, Natal, RN, Brasil}

\affiliation[e]{Millennium Institute for Subatomic Physics at High-Energy Frontier (SAPHIR), Fernandez Concha 700, Santiago, Chile.}

\affiliation[f]{Departamento de F\'isica, Facultad de Ciencias, Universidad de La Serena, Avenida Cisternas 1200, La Serena, Chile.}

\emailAdd{j.echeverriapuentes@uandresbello.edu}
\emailAdd{patricioescalona96@gmail.com}
\emailAdd{farinaldo.queiroz@ufrn.br}
\emailAdd{david.suarezr@udea.edu.co}

\abstract{We will witness enormous progress in the experimental sensitivity to charged-lepton-violation processes in the near future. New physics signals of charged lepton violation might be around the corner without conflicting with existing astrophysical and accelerator bounds. In this work, we explore the possibility of having a dark matter particle as a source for $\mu\to e \gamma$, $\mu \to 3e$, and $\mu\to e$ conversion in nuclei. After computing the dark matter relic density and dark matter-nucleon scattering cross section, we outline the region of parameter space where one can simultaneously accommodate a dark matter fermion in agreement with existing collider and direct detection bounds, and positive signals in charged lepton violation observables.
}

\begin{document}
\maketitle
\newpage
\input{sections/introduction}

\input{sections/model}

\input{sections/dm}
\input{sections/lfv}
\input{sections/discussion}

\input{sections/conclusion}
\input{sections/acknowledgments}
\appendix
\input{sections/appendix}
\bibliographystyle{JHEPfixed.bst}
\bibliography{references}
\end{document}

%% file: sections/introduction.tex
\section{Introduction}
\label{sec:intro}

The Standard Model (SM) provides an excellent description of particle physics phenomena, yet it leaves several established observations unexplained. Two of the most important SM shortcomings that point beyond the Standard Model (BSM) physics are neutrino oscillations and dark matter (DM). The experimental observation of neutrino masses and oscillations~\cite{deSalas:2020pgw,Capozzi:2025wyn,Esteban:2024eli,ParticleDataGroup:2024cfk} has demonstrated that non-SM interactions are needed to generate the neutrino masses, at the same time, which concludes that lepton flavor is not a symmetry of nature. Although there is no experimental confirmation of lepton flavor violation (LFV) processes, intense searches in recent decades keep the door open for their experimental observation and make LFV a sensitive probe of new physics. In parallel, astrophysical and cosmological observations provide overwhelming evidence for the existence of DM~\cite{1980ApJ...238..471R,Planck:2018vyg} and its cosmological behavior, which consolidate the idea that DM is not a SM particle. A well-motivated framework is the Weakly Interacting Massive Particle (WIMP) paradigm, in which DM is thermally produced in the early Universe~\cite{Arcadi:2024ukq,Bertone:2016nfn,deSwart:2017heh}. Experimental efforts probe DM through direct detection~\cite{Goodman:1984dc,Drukier:1986tm,Fan:2010gt}, indirect detection~\cite{HESS:2022ygk,Angel:2025xwb,Angel:2023rdd}, and collider searches~\cite{CMS:2024zqs,Giagu:2019fmp}.

From the perspective of DM phenomenology, simplified models provide an efficient framework for organizing searches in direct detection, indirect detection, and collider experiments~\cite{DiFranzo:2013vra, Berlin:2014tja, BAEK201628, Bell:2016uhg, ElHedri:2017nny, Jacques:2015zha, PhysRevD.91.095020, Backovic:2015soa, Bell_2016, Brennan:2016xjh,Arcadi:2024ukq}. Here, the DM candidate is typically a singlet or weak-isospin multiplet that interacts with the SM via a mediator. While these models are useful for signal-driven studies, they lack a connection to a broader theoretical framework. In contrast, ultraviolet-complete models provide a more complete structure, linking DM to other phenomena such as flavor physics. Representative examples include supersymmetric models~\cite{Griest:2000kj,Chun:2016cnm} and grand unified theories~\cite{Ellis:1988bf,Kadastik:2009cu,Arbelaez:2015ila}. In this spirit, BSM scenarios that explain both phenomena by generating neutrino masses via new LFV interactions with stable particles that cannot have direct interaction with the SM particles are very attractive~\cite{Longas:2023bvq,Restrepo:2021kpq,Agudelo:2024luc,Abada:2014kba,Toma:2013zsa, Lindner:2016bgg}.

In this work, we focus on a class of extensions of the SM model with an extended gauge symmetry to $SU(3)_C\otimes SU(3)_L\otimes U(1)_N$, namely the 331 models~\cite{Mizukoshi:2010ky}. Specifically, we consider the 331LHN realization~\cite{Long:1995ctv,Long:1996rfd}. In this framework, the leptonic sector is enlarged so that each lepton family contains, in addition to the SM neutrino and charged lepton, a heavy neutral fermion. The lightest of these states, the third component of an $SU(3)_L$ lepton triplet, is a viable WIMP candidate~\cite{Mizukoshi:2010ky}, while the extended gauge sector introduces new charged bosons that connect the heavy neutral fermions to the SM charged leptons. As a result, the same gauge interaction that governs the fermionic DM candidate also induces charged LFV. This makes the model particularly predictive, connecting DM phenomenology, LFV observables, and collider signatures ~\cite{Arcadi:2017xbo}.

The 331 framework is also theoretically well motivated. Here, several open questions are addressed, like the flavor structure~\cite{Feruglio:2015jfa,Novichkov:2021evw} and the number of fermion generations~\cite{Pisano:1996ht} since gauge anomaly cancellation requires the number of generations to be a multiple of three. At the same time, asymptotic freedom restricts the number of generations to fewer than five~\cite{Dias:2004dc,Dias:2004wk}, thereby predicting three generations. This class of models is currently under active study~\cite{Aguilar:2025grh,Escalona:2025rxu,Oliveira:2025kfg,Escalona:2025jla,Doff:2026kcs,Kannike:2025qru,Nevzorov:2025ido,Rehman:2025urc}.

The connection between fermionic DM and LFV in the 331LHN model has been explored previously~\cite{Arcadi:2017xbo}. In particular, earlier studies showed that the same heavy neutral fermions that participate in the DM sector can induce the radiative decay $\mu\to e\gamma$, and that the interplay with direct detection and collider limits already imposes significant restrictions on the viable parameter space. Nevertheless, the LFV phenomenology of the model warrants a more thorough reassessment. Beyond the dipole contribution relevant for $\mu\to e\gamma$, the processes $\mu\to 3e$ and coherent $\mu$--$e$ conversion in nuclei receive additional contributions from the non-dipole photon term, $Z^\prime$ penguins, and fermionic box diagrams. These contributions can modify the hierarchy among LFV observables and are therefore essential for a faithful interpretation of the model in light of present and upcoming experiments.

This question is timely given the rapid experimental progress. Current and upcoming searches for $\mu\to e\gamma$, $\mu\to 3e$, and $\mu$--$e$ conversion are significantly improving the sensitivity to charged LFV as can be seen in \autoref{tab:signal}. In parallel, direct detection experiments impose stringent limits on the spin--independent DM--nucleon scattering cross section~\cite{LZ:2024zvo,XENON:2025vwd,PandaX:2024qfu,XLZD:2024nsu,DARWIN:2016hyl}. For DM masses above $\mathcal{O}(10)\,\mathrm{GeV}$, the leading constraints are set by liquid-xenon time-projection chamber experiments, with the most stringent current bound set by the LZ~\cite{LZ:2024zvo} experiment, closely followed by XENONnT~\cite{XENON:2025vwd} and PandaX-4T~\cite{PandaX:2024qfu}. Future detectors such as XLZD~\cite{XLZD:2024nsu} and DARWIN~\cite{DARWIN:2016hyl} will further extend the reach by up to one order of magnitude. Together, these complementary probes enable a comprehensive assessment of the viable parameter space and highlight the scenario's near-term testability. Additionally, colliders can probe new neutral gauge bosons through dilepton resonance searches. Such searches place complementary constraints on the mass of the $Z^\prime$ boson, and consequently on the energy scale of the enlarged $SU(3)_L$ symmetry.

The goal of this work is to revisit the 331LHN model in light of current experimental constraints, with particular emphasis on the interplay between fermionic DM and charged LFV. We update the bounds arising from relic abundance, spin-independent direct detection, and collider searches, and we extend the LFV analysis by including, in addition to the dipole contribution, the non-dipole photon term, the $Z^\prime$ penguin, and the relevant fermionic box diagrams. This allows us to determine more precisely how the viable parameter space is organized once all constraints are imposed simultaneously. We show that the region compatible with DM viability remains predominantly dipole-dominated, in agreement with previous approximations. In contrast, outside this regime, additional contributions can substantially modify the hierarchy among LFV observables, with $\mu$--$e$ conversion emerging as the most distinctive future probe of the model.

The paper is organized as follows. In \autoref{sec:model}, we summarize the structure of the model and the ingredients relevant for DM and LFV. In \autoref{sec:dm} we discuss the fermionic DM phenomenology, including relic abundance and direct detection. In \autoref{Lepton_Flavor_Violation} we present the LFV analysis for $\mu\to e\gamma$, $\mu\to 3e$, and $\mu$--$e$ conversion. In \autoref{discussion}, we combine DM, LFV, and collider constraints and identify the viable parameter space of the model. Finally, we present our conclusions in \autoref{conclusions}.

\begin{table}[t]
\centering
\renewcommand{\arraystretch}{1.2}
\setlength{\tabcolsep}{8pt}
\begin{tabular}{lcc}
\toprule
\textbf{Process} & \textbf{Current limit} & \textbf{Prospect} \\
\midrule
$\mu \to e \gamma$ 
& $\operatorname{Br}<1.5 \times 10^{-13}$~\cite{MEGII:2025gzr} 
& $\operatorname{Br}<6 \times 10^{-14}$~\cite{MEGII:2025gzr} \\

$\mu \to e e \bar{e}$ 
& $\operatorname{Br}<1 \times 10^{-12}$~\cite{SINDRUM:1987nra} 
& $\operatorname{Br}<1 \times 10^{-16}$~\cite{Miscetti:2025uxk}\\

$\mu$--$e$ conversion (Au) 
& $\mathrm{CR}(\mu\text{--}e,\mathrm{Au})<7\times10^{-13}$~\cite{SINDRUMII:2006dvw} 
& --- \\

$\mu$--$e$ conversion (Al) 
& --- 
& $\mathrm{CR}(\mu\text{--}e,\mathrm{Al})<2\times10^{-17}$~\cite{COMET:2018auw} \\
\bottomrule
\end{tabular}
\caption{Current $90\%$ C.L. upper bounds and future sensitivities for representative charged lepton flavor violating muon processes. Status of the experiments can be found in Ref.~\cite{COMET:2025sdw}.}
\label{tab:signal}
\end{table}

%% file: sections/model.tex
\section{Model Overview}
\label{sec:model}
We consider a $SU(3)_C \otimes SU(3)_L \otimes U(1)_N$ gauge theory. In this family of models, the generalized Gell-Mann-Nishijima is given by
\begin{equation}
\frac{\hat{Q}}{e}=\frac{1}{2}(\lambda_3 + \beta \lambda_8) +N I,
\end{equation}
where $e$ is the electric charge of the electron, $\lambda_{3,8}$ are the diagonal Gell-Mann matrices, $N$ is the $U(1)_N$ charge, $I$ is the identity operator, and $\beta$ is a model-dependent parameter. In this work, we consider the 331LHN model, which features $\beta=-1/\sqrt{3}$.

The $SU(3)_L$ gauge bosons are denoted by $W^a$ with $a=1,\dots,8$, and the $U(1)_N$ gauge boson is denoted by $W^N$. These gauge bosons linearly mix to form the following spectrum:
\begin{align}
W^{\pm}_{\mu}= & \frac{1}{\sqrt{2}}(W^1_{\mu}\mp{i}W^2_{\mu})\,,\\
U^0_{\mu}= & \frac{1}{\sqrt{2}}(W^4_{\mu}-iW^5_{\mu})\,,\\
W^{\pm\prime}_{\mu}= & \frac{1}{\sqrt{2}}(W^6_{\mu}\pm{i}W^7_{\mu})\,.\\
B_{\mu}= & -\frac{t_W}{\sqrt{3}}W^8_{\mu}+\sqrt{1-\frac{t^2_{W}}{3}}W^N_{\mu}\,,\\
Z'_{\mu}= & \sqrt{1-\frac{t^2_{W}}{3}}W^8_{\mu}+\frac{t_W}{\sqrt{3}}W^N_{\mu}\,,
\end{align}
where $t_W$ is the tangent of the Weinberg angle. $W^\pm$ are the SM charged gauge bosons, while $W^{\pm\prime }$ are new, heavier ones. $U^0$ is a non-Hermitian neutral boson, and $Z^\prime$ is a new Hermitian neutral gauge boson.  Moreover, $B_{\mu}$ is the hypercharge field of the SM, so $W_\mu^3$ and $B^\mu$ form the photon and the $Z$ boson in the usual way: $A_\mu = s_W W_\mu^3 +c_W B_\mu$ and $Z_\mu = c_W W_\mu^3 -s_W B_\mu$.

In this model, we arrange SM leptons in the fundamental representation of $SU(3)_L$, together with heavy neutral left-handed fermions~\cite{Mizukoshi:2010ky,Arcadi:2017xbo}:
\begin{equation}
\psi_{aL}=\begin{pmatrix}
\nu_a\\
e_a\\
N_a
\end{pmatrix}_L\sim(\mathbf{1},\mathbf{3},-1/3)\,,
\end{equation}
where $a=1,2,3$ is a family index, and the terms in parentheses denote the $SU(3)_C \otimes SU(3)_L \otimes U(1)_N$ quantum numbers, respectively. The right-handed leptons are $SU(3)_L$ singlets:
\begin{align}
e_{aR}\sim(\mathbf{1},\mathbf{1},-1)\,, \qquad N_{aR}\sim(\mathbf{1},\mathbf{1},0)\,.
\end{align}
Assuming tree-level lepton number conservation, a Majorana mass term for $N_{aR}$ is forbidden. As a result, the new heavy neutral leptons have Dirac masses. The lightest of these three Dirac fermions constitutes the DM candidate.\footnote{A Majorana fermion in this framework would lead to a qualitatively different DM phenomenology.}. 

In the quark sector, we arrange the left-handed SM quarks in the fundamental representation of $SU(3)_L$, along with exotic quarks $U$ and $D$ with usual electric charges $+2/3$ and $-1/3$. Anomaly cancellation requires one quark generation to transform differently from the other two. We impose that the first two families transform as $SU(3)_L$ antitriplets, while the third quark family transforms as a triplet:
\begin{equation}
Q_{\alpha{L}}=\begin{pmatrix}
d_{\alpha}\\
-u_{\alpha}\\
D_{\alpha}
\end{pmatrix}_L\sim(\mathbf{3},\mathbf{3}^*,0)\,,    \qquad
Q_{3{L}}=\begin{pmatrix}
u_3\\
d_3\\
U_3
\end{pmatrix}_L\sim(\mathbf{3},\mathbf{3},1/3)\,. 
\end{equation}
where $\alpha=1,2$ labels the first two families. The right-handed quarks transform trivially under $SU(3)_L$: 
\begin{align}
u_{\alpha{R}}\sim & (\mathbf{3}, \mathbf{1},2/3)\,, \qquad
d_{\alpha{R}}\sim  (\mathbf{3},\mathbf{1},-1/3)\,, \qquad
D_{\alpha{R}}\sim  (\mathbf{3},\mathbf{1},-1/3)\,, \nonumber\\
u_{3R}\sim & (\mathbf{3},\mathbf{1},2/3)\,, \qquad
d_{3R}\sim  (\mathbf{3},\mathbf{1},-1/3)\,, \, \qquad
U_{3R}\sim  (\mathbf{3},\mathbf{1},2/3)\,.
\end{align}

The scalar sector consists of three triplets\footnote{Models with fewer scalar triplets are excluded~\cite{Ferreira:2011hm,Phong:2013cfa}.}. To generate mass terms for quarks, charged leptons, and gauge bosons, the scalar fields must transform as
\begin{align}
\eta=  \begin{pmatrix}
\eta^0\\
\eta^-\\
\eta^{0'}
\end{pmatrix}\sim(\mathbf{1},\mathbf{3},-1/3)\,, \quad
\rho=  \begin{pmatrix}
\rho^+\\
\rho^0\\
\rho^{+'}
\end{pmatrix}\sim(\mathbf{1},\mathbf{3},2/3)\,, \quad 
\chi= & \begin{pmatrix}
\chi^0\\
\chi^-\\
\chi^{0'}
\end{pmatrix}\sim(\mathbf{1},\mathbf{3},-1/3)\,, \quad
\end{align}
The neutral components of these triplets that acquire VEVs are $\eta^0,\rho^0$ and $\chi^{0\prime}$, while the remaining two neutral components are inert. 

To stabilize the DM, an R-parity-like symmetry $P=(-1)^{3(B-L)+2s}$ is imposed on the Lagrangian. Explicitly, the following fields are odd under this $Z^2$ symmetry:
\begin{equation}
\{\, N_a,\; D_{\alpha},\; U_3,\; \rho^{+'},\; \eta^{0'},\; \chi^0,\; W',\; U^0 \,\}
\end{equation}
while the remaining fields (including all SM fields) transform trivially. The lightest odd neutral state constitutes the DM candidate. Note that states with odd R-parity have non-trivial lepton number. For instance, the new quarks $U$ and $D$ carry lepton and baryon number\footnote{Particles with this property are called \textit{leptoquarks}.}.

The spontaneous symmetry breaking occurs in two steps.
First, when $\chi^{0'}$ acquires a VEV, the enlarged group is broken to the SM electroweak group. Then, the VEVs of $\eta^0$ and $\rho^0$ break the electroweak theory to QED. Schematically,
\begin{equation}
SU(3)_L\otimes{U}(1)_N\xrightarrow{\langle\chi\rangle}SU(2)_L\otimes{U}(1)_Y\xrightarrow{\langle\rho\rangle,\langle\eta\rangle}U(1)_Q \,.   
\end{equation}
For simplicity, we consider $v_{\eta}=v_{\rho}=v$ such that $\sqrt{2}v\equiv \sqrt{v^2_{\eta}+v^2_{\rho}}={246}\text{ GeV}$. 

The scalar potential is \ footnote {Lepton number conservation forbids additional terms.}
\begin{align}
V(\eta,\rho,\chi) =\;&
\mu^2_{\chi}\,\chi^{\dagger}\chi
+\mu^2_{\eta}\,\eta^{\dagger}\eta
+\mu^2_{\rho}\,\rho^{\dagger}\rho
\nonumber\\[4pt]
&+ \lambda_1 (\chi^{\dagger}\chi)^2
+ \lambda_2 (\eta^{\dagger}\eta)^2
+ \lambda_3 (\rho^{\dagger}\rho)^2
\nonumber\\[6pt]
&+ \lambda_4 (\chi^{\dagger}\chi)(\eta^{\dagger}\eta)
+ \lambda_5 (\chi^{\dagger}\chi)(\rho^{\dagger}\rho)
+ \lambda_6 (\eta^{\dagger}\eta)(\rho^{\dagger}\rho)
\nonumber\\[6pt]
&+ \lambda_7 (\chi^{\dagger}\eta)(\eta^{\dagger}\chi)
+ \lambda_8 (\chi^{\dagger}\rho)(\rho^{\dagger}\chi)
+ \lambda_9 (\eta^{\dagger}\rho)(\rho^{\dagger}\eta)
\nonumber\\[6pt]
&- \frac{f}{\sqrt{2}}
\left(
\epsilon^{ijk}\,\eta_i \rho_j \chi_k
+ \text{h.c.}
\right)\,.
\end{align}

We now consider the Yukawa Lagrangian. For charged leptons is
\begin{equation}
\mathcal{L}^l_Y=h^l_{ab}\overline{\psi}_{aL}\rho{e}_{bR}\,,    
\end{equation}
while for the heavy neutral leptons, it corresponds to~\footnote{Lepton number conservation forbids additional terms.}
\begin{equation}
\mathcal{L}^N_Y=  g'_{ab}\overline{\psi}_{aL}\chi{N}_{bR}\,.   
\end{equation}
Additionally, the Yukawa terms for quarks read
\begin{align}
\mathcal{L}^q_Y= & h^u_{\alpha{a}}\overline{Q}_{\alpha{L}}\rho^*u_{aR}+h^d_{\alpha{a}}\overline{Q}_{\alpha{L}}\eta^*d_{aR}+h^U\overline{Q}_{3L}\chi{U}_{3R}\nonumber\\
& +h^d_{3a}\overline{Q}_{3L}\rho{d}_{aR}+h^u_{3a}\overline{Q}_{3L}\eta{u}_{aR}+h^d_{\alpha\beta}\overline{Q}_{\alpha{L}}\chi^*d_{\beta{R}}+\mathrm{h.c}\,,
\end{align}

In the following, we list the resulting mass spectrum. The squared masses of the new gauge bosons $Z'$, $W'$, and $U^0$ are
\begin{align}
M^2_{Z'}= & \frac{g^2_L}{4(3-4s^2_W)}\left[4c^2_Wv^2_{\chi'}+\frac{v^2}{c^2_W}+\frac{v^2(1-2s^2_W)^2}{c^2_W}\right]\,,  \label{eq:mzmass}\\
M^2_{W'}=&M^2_{U^0}=\frac{g^2_L}{4}\left(v^2+v^2_{\chi'}\right)\,.\label{eq:mwmass}
\end{align}
If $v_{\chi'}\gg{v}$, the masses of the new gauge bosons are approximately $M_{Z'}\cong{0.4}v_{\chi'}$ and $M_{W'}\cong{0.32}v_{\chi'}$. In this regime, the large mass difference between $Z$ and $Z^{\prime}$ suppresses their mixing. Thus, we neglect it.

The  charged lepton mass matrix is given by
\begin{equation}
M^l=h^l\frac{v_{\rho}}{\sqrt{2}}\,,
\end{equation}
and the heavy neutral lepton mass matrix is
\begin{equation}\label{fermionmass}
M^N=\frac{1}{\sqrt{2}}g'v_{\chi'}\,.    
\end{equation}
A biunitary transformation diagonalizes the mass matrix
\begin{equation}
UM^NV^{\dagger}=M^N_{\text{diag}}    
\end{equation}
where $V$ is the identity for simplicity, the lightest mass eigenstate $N_1$ is the dark matter candidate.
If we assume a diagonal charged lepton mass matrix, the $U$ matrix quantifies the mixing between heavy neutral fermions and charged leptons, which is crucial for LFV muon decays and transitions.

For the exotic and SM quarks, the corresponding mass matrices are
\begin{align}
M^u= & \frac{1}{\sqrt{2}}\begin{pmatrix}
-v_{\rho}h^u_{11} & -v_{\rho}h^u_{12} & -v_{\rho}h^u_{13} \\
-v_{\rho}h^u_{21} & -v_{\rho}h^u_{22} & -v_{\rho}h^u_{23} \\
v_{\eta}h^u_{31}  &  v_{\eta}h^u_{32} &  v_{\eta}h^u_{33}
\end{pmatrix}\,,\\
M^d= & \frac{1}{\sqrt{2}}\begin{pmatrix}
v_{\eta}h^d_{11} & v_{\eta}h^d_{12} & v_{\eta}h^u_{13} \\
v_{\eta}h^d_{21} & v_{\eta}h^d_{22} & v_{\eta}h^u_{23} \\
v_{\rho}h^u_{31} & v_{\rho}h^u_{32} & v_{\rho}h^u_{33}
\end{pmatrix}\,,\\
M^U= & \frac{1}{\sqrt{2}}v_{\chi'}h^U\,,\\
M^D= & \frac{1}{\sqrt{2}}\begin{pmatrix}
v_{\chi'}h^D_{11} & v_{\chi'}h^D_{12} \\
v_{\chi'}h^D_{21} & v_{\chi'}h^D_{22}
\end{pmatrix}\,.
\end{align}

We now turn to the scalar spectrum. For simplicity, we assume $\lambda_1=1 ,\lambda_4=\lambda_5=\frac{1}{4}$, $f=\frac{v_{\chi'}}{2}$. out of the 18 degrees of freedom of the triplets, 10 correspond to physical states: three CP-even eigenstates $S_1$, $S_2$, and $H$ with masses given by
\begin{align}
M^2_{S_1}= & \,\frac{1}{2}\left(\frac{v^2}{2}+4v^2_{\chi'}\right)\,,\\
M^2_{S_2}\approx & \,\frac{1}{2}\left(v^2_{\chi'}+2v^2(\lambda_2+\lambda_3-\lambda_6)\right)\,,\\
M^2_{H}\approx \, & v^2(\lambda_2+\lambda_3+\lambda_6)\,,
\end{align}
one CP-odd eigenstate with mass given by
\begin{equation}
M^2_{P_1}=\frac{1}{2}\left(v^2_{\chi'}+\frac{v^2}{2}\right),
\end{equation}
a complex scalar $\phi_1$ with mass given by
\begin{equation}
M^2_{\phi_1}=\frac{1}{2}\left(\lambda_8+\frac{1}{2}\right)(v^2+v^2_{\chi'})
\end{equation}
and finally, two charged $h_1^-$ and $h_2^-$ with  masses
\begin{align}
M^2_{h^-_1} = & \frac{1}{2}\left(\lambda_8+\frac{1}{2}\right)(v^2+v^2_{\chi'})\,,\\
M^2_{h^-_2} = & \frac{v^2_{\chi'}}{2}+\lambda_9v^2
\end{align}
In addition, the scalar spectrum comprises eight Goldstone bosons that become the longitudinal components of the massive gauge bosons after SSB.

For the new gauge bosons $W'$ and $Z'$, the relevant fermionic currents are
\begin{equation}
\label{eq:fermion_interactions}
    \begin{aligned}
    \mathcal{L}= & -\frac{g}{\sqrt{2}}[\overline{N}_{aL}\gamma^{\mu}l_{aL}W^{'+}_{\mu}+\overline{U}_{3L}\gamma^{\mu}d_{3L}W^{'+}_{\mu}+\overline{u}_{a L}\gamma^{\mu}D_{\alpha{L}}W^{'+}_{\mu}]\\
    &-\frac{g}{2\cos\theta_W}\sum_f\left[\overline{f}\gamma^{\mu}(g_V+g_A\gamma^5)fZ'_{\mu}\right]\,.
    \end{aligned}
\end{equation}

%% file: sections/dm.tex
\section{Dark Matter}
\label{sec:dm}
As stated in \autoref{sec:model}, the model features an $R$-parity-like symmetry, and there are three neutral particles with odd parity: the fermions $N_{1,2,3}$, the non-Hermitian gauge boson $U^0$, and the complex scalar $\phi_1$. The authors study scalar DM in Ref.~\cite{Mizukoshi:2010ky}. When $m_{N_1}>m_{W^\prime}=m_{U^0}$, the $U^0$ relic density is strongly suppressed, $\Omega h^2\sim10^{-6}$. Therefore, we focus on the case in which a fermion is the lightest of these states, so it constitutes the DM candidate. Without loss of generality, we take $N_1$ to be the DM candidate\footnote{More precisely, a linear combination of $N_{1,2,3}$ after mass mixing.} From \eqref{fermionmass}, the entries of the $g^\prime$ matrix must be such that the eigenvalues of the mass matrix $M^N$ satisfy $m_{N_1}<m_{W^\prime}\approx{0.32} v_{\chi^\prime}$, ensuring its stability.

The relic abundance and spin-independent DM-nucleon scattering then give the relevant DM phenomenology. Although indirect detection searches are promising~\cite{CTAConsortium:2017dvg}, such signals are beyond the scope of this work. The mixing of $N_1$ and charged leptons has a negligible effect on the DM phenomenology~\cite {Profumo:2013sca}. 

We consider the thermal decoupling of $N_1$, driven by the pair annihilation and coannihilation with $N_2$ and $N_3$. Note that the kinematic constraints presented earlier forbid mediators from appearing in the final states, so the secluded DM scenario is not plausible. In the following, we discuss in some detail the freeze-out mechanism in the $\Lambda_{\rm CDM}$ cosmology\footnote{Non-standard cosmology effects can modify the resulting relic abundance~\cite {Gelmini:2006mr,Gelmini:2006pw,Arcadi:2011ev,Baer:2014eja}.}.

\subsection{Relic Abundance}
\begin{figure}[]
  \centering
  \subfigure[]{\includegraphics[width=.325\linewidth]{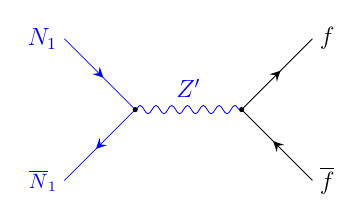}}
  \subfigure[]{\includegraphics[width=.325\linewidth]{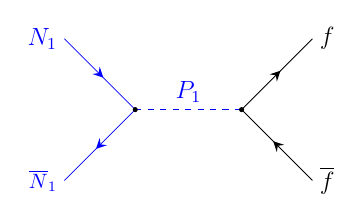}}
  \subfigure[]{\includegraphics[width=.325\linewidth]{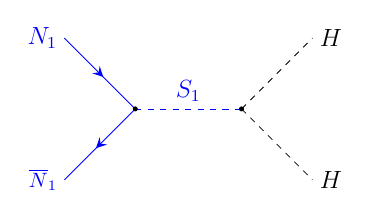}}
  \subfigure[]
  {\includegraphics[width=.325\linewidth]{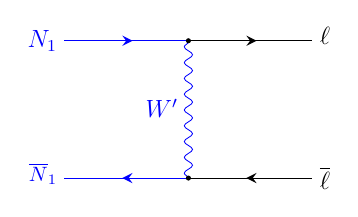}}
  \subfigure[]
  {\includegraphics[width=.325\linewidth]{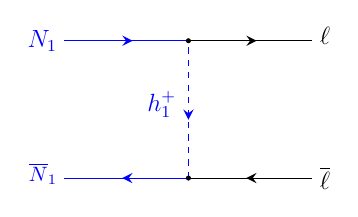}}
  \caption{Feynman diagrams relevant for the DM relic abundance. Panels (a) and (b) show the dominant $s$-channel annihilation of the fermionic DM candidate $N_1$ into SM fermions through $Z^\prime$ and $P_1$, respectively. Panel (c) shows the analogous contribution mediated by the CP-even scalar $S_1$. Panels (d) and (e) display the $t$-channel annihilation into charged leptons mediated by $W^\prime$ and $h_1^\pm$.}\label{fig:DM_Anihilation}
\end{figure}

The diagrams for DM annihilation into fermions, mediated by the $Z^\prime$ boson and the pseudoscalar $P_1$, are depicted in \ref{fig:DM_Anihilation}(a) and \ref{fig:DM_Anihilation}(b), and the corresponding cross-sections can be found in Eqs. (15-16) from Ref~\cite{Arcadi:2017xbo}. Annihilation via the CP-even scalar $S_1$ (Fig. \ref{fig:DM_Anihilation}(c)) is subdominant because this particle is twice as heavy as $P_1$ in our simplified scalar sector, and much heavier than $Z^\prime$. Processes in the t-channel mediated by $W^\prime$ or a charged scalar (Figs \ref{fig:DM_Anihilation}(d) and \ref{fig:DM_Anihilation}(e)) also contribute negligibly to DM relic abundance. Nevertheless, we use micrOmegas~\cite{Belanger:2001fz,Belanger:2006is,Belanger:2010gh,Belanger:2013oya,Belanger:2014vza,Belyaev:2012qa,Alguero:2023zol} to numerically solve the Boltzmann equation considering all relevant channels, including coannihilation with heavier fermions $N_2$ and $N_3$. 

\begin{figure}
    \centering
    \includegraphics[width=0.75\linewidth]{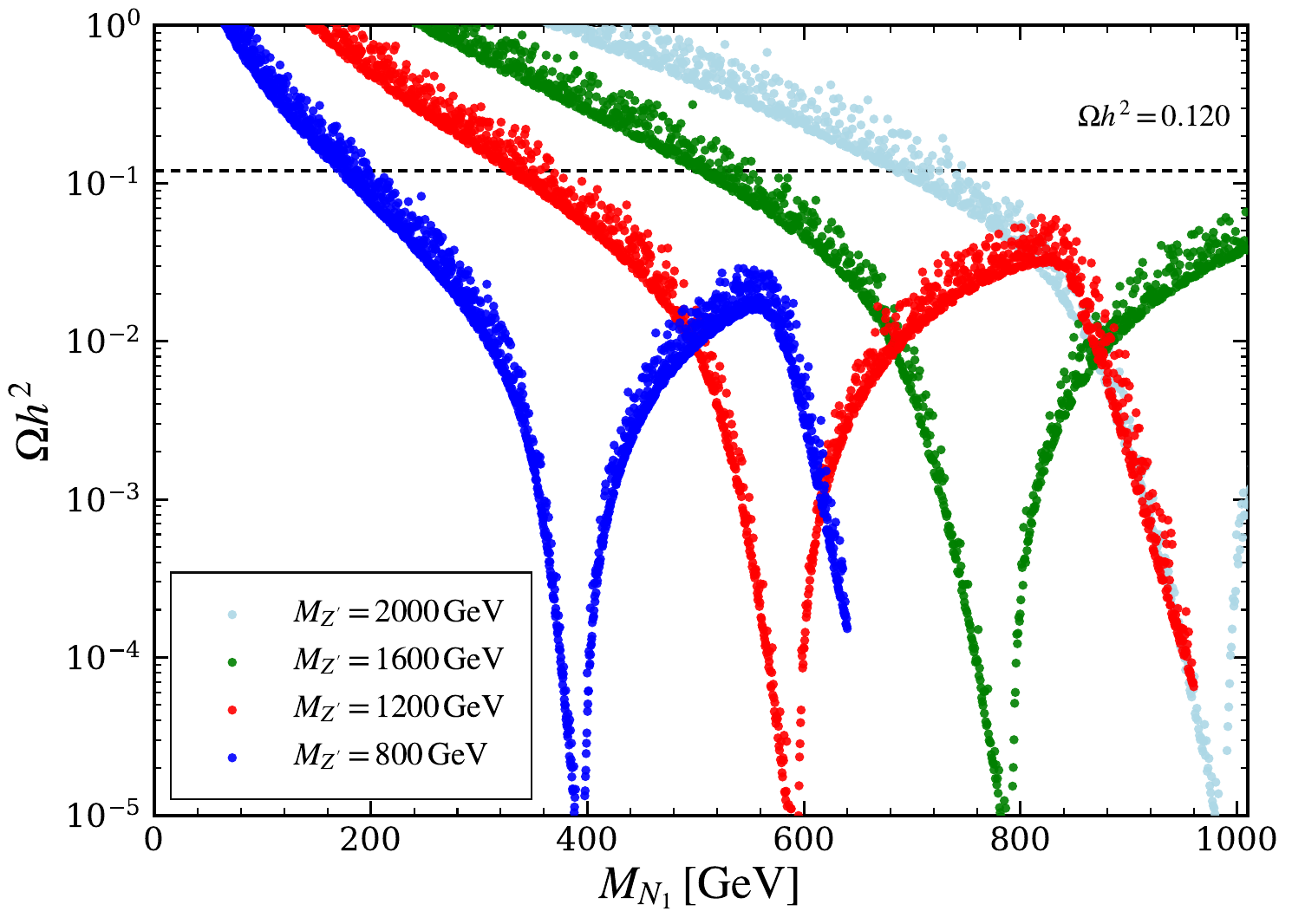}
     \caption{Relic abundance of the fermionic dark matter candidate $N_1$ as a function of $M_{N_1}$ for representative values of $M_{Z'}$. The horizontal dashed line indicates the observed value $\Omega h^2=0.120$ \cite{Planck:2018vyg}. The narrow dips correspond to resonant enhancement of the annihilation cross section, while the intersections with the Planck value define the viable DM mass windows for each $Z'$ mass.}
    \label{fig:darkmatter}
\end{figure}

We show our findings in~\autoref{fig:darkmatter}, where we plot the relic abundance of the DM candidate as a function of the mass of the DM candidate, for different mediator masses, and the horizontal line corresponds to the value of $\Omega{h}^2$ reported by Planck~\cite{Planck:2018vyg}. The major impact on the relic density curve is due to the $Z^\prime$ resonance: as we approach it, the DM annihilation cross section increases, and the relic abundance decreases. The mediating pseudoscalar, heavier than $Z^\prime$, also impacts the relic density from its resonance. To illustrate the effect of coannihilation with nearly degenerate fermions $N_2$ and $N_3$, we consider those states to be randomly at most $15\%$ heavier than $N_1$. The coannihilation effect is the origin of the apparent thickness of the relic abundance curves.

\subsection{Direct Detection}
\begin{figure}
    \centering
    \includegraphics[width=0.75\linewidth]{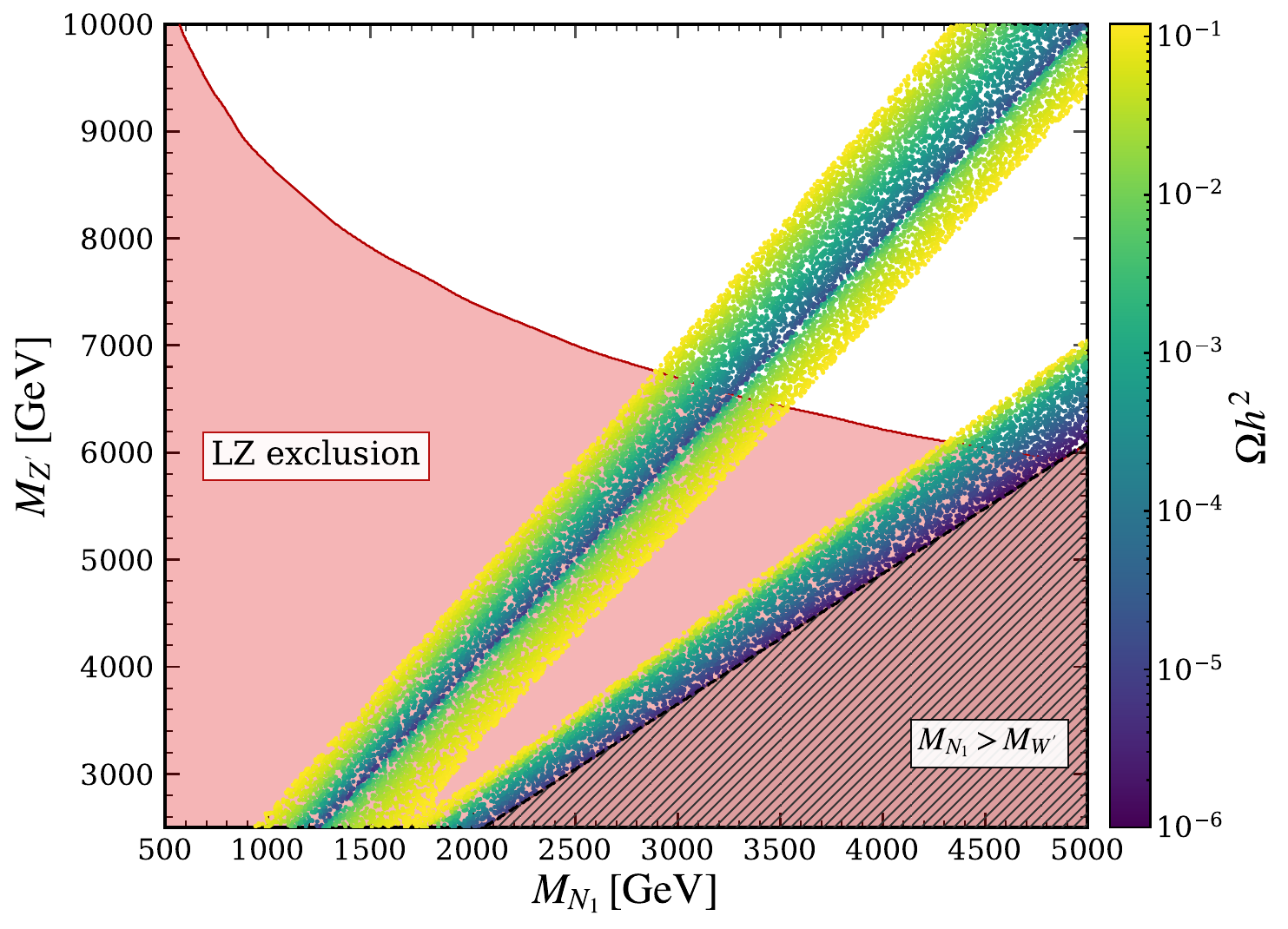}
\caption{Spin-independent direct-detection constraints in the $M_{Z'}$--$M_{N_1}$ plane. LZ excludes the red shaded region \cite{LZ:2024zvo}, while the hatched region with $M_{N_1}>M_{W'}$ corresponds to an unstable DM candidate. The colored points show the predicted relic abundance $\Omega h^2$, and the diagonal bands identify the resonant regions where the model yields the observed abundance or sub-abundant DM.}
    \label{fig:DD}
\end{figure}

In this model, the $Z^\prime$ exchange in the t-channel mediates the relevant DM–nucleon scattering process. The pseudoscalar contributes negligibly to direct detection~\cite{Arcadi:2017wqi}. The spin-independent cross-section is
 \begin{equation}
\sigma_{\text{SI}}\approx\frac{\mu^2_{Nn}}{\pi}\left[\frac{Zf_p+(A-Z)f_n}{A}\right]^2     
 \end{equation}
 where $\mu^2_{Nn}$ is the reduced mass of the DM-nucleon system, $f_p=1/{M^2_{Z^\prime}}(2g^\prime_{NuV}+g^\prime_{NdV})$ and $f_n=1/{M^2_{Z^\prime}}(g^\prime_{NuV}+2g^\prime_{NdV})$. The specific values of $g'_{NuV}$ and 
 $g'_{NdV}$ can be found in Ref.~\cite{Arcadi:2017xbo}. We use micrOmegas~\cite{Belanger:2001fz,Belanger:2006is,Belanger:2010gh,Belanger:2013oya,Belanger:2014vza,Belyaev:2012qa,Alguero:2023zol} to compute the relevant cross section for scattering of DM off nuclei. We plot our findings in \autoref{fig:DD}, where we show the bounds from the LZ direct detection experiment in the $M_{Z^{\prime}}$-$M_{N_1}$ plane. LZ excludes the red region because the spin-independent cross-section exceeds current upper bounds. The region filled with black lines corresponds to the $N_1$ instability region. The colored bands correspond to values with the correct or sub-abundant DM relic density.

%% file: sections/lfv.tex
\section{Lepton Flavor Violation}
\label{Lepton_Flavor_Violation}

LFV is among the most promising probes of new physics. In particular, we focus on the searches for the muon decays $\mu \to e\gamma$ and $\mu \to ee\bar{e}$, as well as on muon-to-electron conversion in nuclei\footnote{We consider muon decays because of the high muon production rate and their clean final states. Processes involving tau leptons have weaker bounds~\cite{Lychkovskiy:2010ue,Dinh:2012bp,Abada:2014kba,He:2014efa}.}. In this model, these processes are especially relevant as they arise from the $W^\prime N\ell$ interaction. Previous studies examined the region dipolar contribution in the region compatible with DM~\cite{Arcadi:2017xbo}. In contrast, this work extends the analysis to include contributions from the $Z'$ penguin and fermionic box diagrams, thereby probing regions beyond the dipole-dominated regime. To compute the diagrams, we follow the compendium of contributions to LFV amplitudes in \cite{Abada:2014kba} in the Feynman gauge and in the limit $m_e, m_\mu \ll M_{N_1}, M_{W^\prime}$.

It is important to consider these three processes at once because each probes a different aspect of the model. The process $\mu \to e\gamma$ probes dipole operators only, $\mu \to ee\bar{e}$ probes non-dipole fermionic corrections, and $\mu$--$e$ conversion includes contributions from the exotic quark sector, whose structure is a distinctive feature of the model.

\subsection{$\mu\to{e}\gamma$}
\begin{figure}
    \centering
    \includegraphics[width=0.4\linewidth]{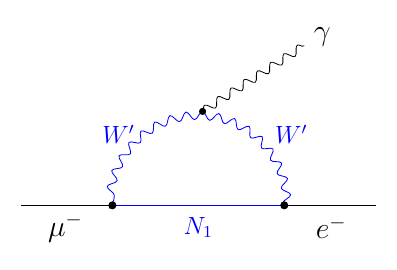}
    \caption{One-loop diagram contributing to $\mu \to e\gamma$, mediated by $N_1$ and $W^\prime$. Diagrams with Goldstone bosons and self-energy corrections over the external legs have also been included.}
    \label{fig:mu_e_gamma}
\end{figure}
This decay is particularly important as it is subject to the most stringent current limits, as shown in Table~\ref{tab:signal}. In this model, the dominant contribution arises from the diagram in Fig.~\ref{fig:mu_e_gamma}, mediated by the virtual exchange of the lightest fermion, $N_1$. We neglect the contributions from the heavier states $N_2$ and $N_3$. This process is mediated exclusively by the dipole operator~\cite{Abada:2014kba,Lavoura:2003xp}. The decay width for $\mu \to e \gamma$ in this model is
\begin{equation}
\label{eq:mutogaamma}
\Gamma(\mu \to e \gamma)= \frac{\alpha_{EM}m_\mu^5}{4}\left(\frac{g^2}{8(4\pi)^2M^2_{W^{'\pm}}}F_2(x)\right)^2\left|U^{N_1 e *} U^{N_1 \mu}\right|^2
\end{equation}
where $x=\frac{M_{N_1}^2}{M_{W'}^2}$, and $F_2(x)$ is the loop function associated with the dipole part of the effective photon interaction. Here, $U^{N_i\ell_j}$ denotes the matrix element mixing the $i$th heavy fermion and the $j$th charged lepton. We provide further details in Appendix~\ref{app:LFV_expressions}. 

\subsection{$\mu\to{3e}$}
\begin{figure}[]
  \centering
  \subfigure[]{\includegraphics[width=.325\linewidth]{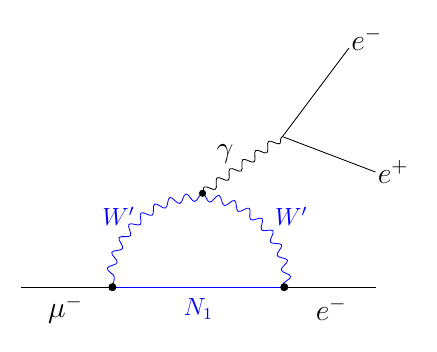}}
  \subfigure[]{\includegraphics[width=.325\linewidth]{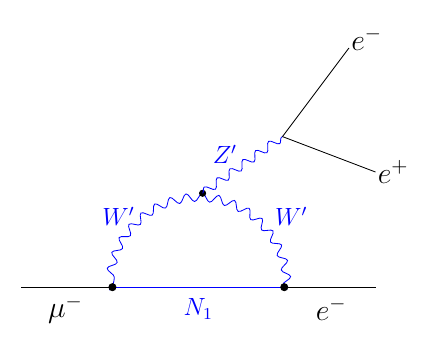}}
  \subfigure[]{\includegraphics[width=.325\linewidth]{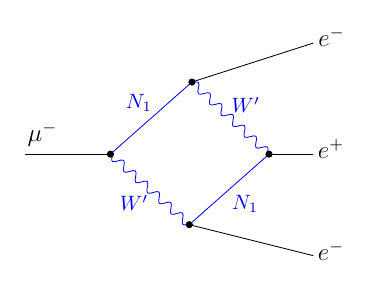}}
  \caption{Representative one-loop contributions to the decay $\mu \to 3e$. Panel (a) shows the off-shell photon contribution, which contains both dipole and non-dipole terms. Panel (b) corresponds to the $Z^\prime$ penguin contribution, while panel (c) shows the leptonic box contribution. Diagrams with Goldstone bosons, permutation, and self-energy corrections over the external legs have also been included.}\label{fig:mu_3e_diagrams}
\end{figure}
The decay $\mu \to ee\bar{e}$ receives contributions from both the dipole and non-dipole parts of the off-shell photon interaction, as well as from leptonic box contributions, as shown in \autoref{fig:mu_3e_diagrams}. The decay width is
\begin{equation}
\label{eq>muto3e}
\begin{aligned}
    \Gamma (\mu \to 3e) =& \frac{m_\mu^5}{512 \pi^3}\Bigg[e^4\left(\frac{16}{3}\log \frac{m_\mu}{m_e}-\frac{22}{3}\right)\left(\frac{g^2}{8(4\pi)^2M^2_{W^{'\pm}}}F_2(x)\right)^2\left|U^{N_1 e *} U^{N_1 \mu}\right|^2\\
    &+\frac{2}{3}\left|\hat A_{LL}^V\right|^2
    +\frac{1}{3}\left|\hat A_{LR}^V\right|^2
    -\frac{8e^2}{3}\text{Re}\left(K_2^R\hat A_{LL}^{V*}\right)
    -\frac{4e^2}{3}\text{Re}\left(K_2^R\hat A_{LR}^{V*}\right)\Bigg],
\end{aligned}    
\end{equation}
with 
\begin{equation}
\begin{aligned}
\hat A_{LL}^V &= \Delta_{Z^\prime L}^{(\ell)}+\Delta_{\rm box}^{(\ell)}+e^2K_1^L,\\
\hat A_{LR}^V &= \Delta_{Z^\prime R}^{(\ell)}+e^2K_1^L.
\end{aligned}
\end{equation}
The first term in \eqref{eq>muto3e} corresponds to the dipolar contribution of the off-shell photon, the second and third terms are leptonic contact contributions, and the final two terms account for the interference between diagrams. Although loop-suppressed, their distinct operator structure allows them to compete with the dipole piece in parts of parameter space. In the dipole-dominated limit, we recover the known linear relation with the $\mu \to e\gamma$ decay width, $\Gamma(\mu \to 3e)\simeq \frac{1}{163}\Gamma(\mu \to e\gamma)$. In the full calculation, however, this relation is no longer exact due to the $Z^\prime$ and leptonic box contributions. The explicit expressions for $\Delta_\gamma$, $\Delta_{Z^\prime}$ and $\Delta_{\rm box}^{(\ell)}$ are collected in Appendix~\ref{app:LFV_expressions}.

\subsection{$\mu-e$ conversion in nuclei}
\begin{figure}[]
  \centering
  \subfigure[]{\includegraphics[width=.325\linewidth]{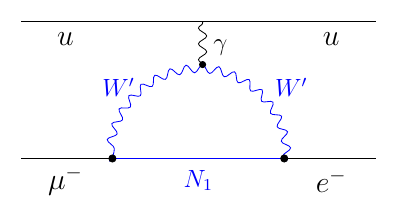}}
  \subfigure[]{\includegraphics[width=.325\linewidth]{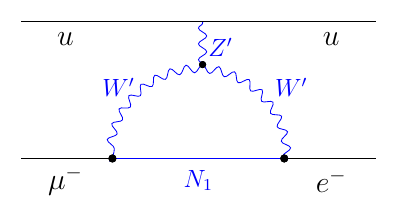}}
  \subfigure[]{\includegraphics[width=.325\linewidth]{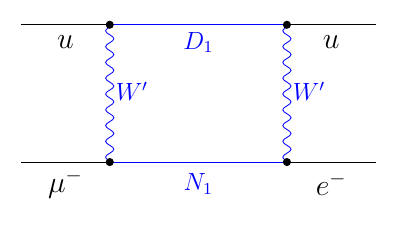}}
  \caption{Contributions to coherent $\mu$--$e$ conversion in nuclei. Panel (a) shows the off-shell photon contribution, panel (b) the $Z^\prime$ penguin contribution, and panel (c) the quark-lepton box diagram involving the exotic quark sector.}\label{fig:mu_e_conv_diagrams}
\end{figure}
Unlike $\mu \to e\gamma$, coherent $\mu$--$e$ conversion in nuclei is sensitive to both the electromagnetic dipole operator and effective four-fermion interactions involving quark vector currents, as depicted in \autoref{fig:mu_e_conv_diagrams}. For this reason, the conversion rate cannot, in general, be inferred from a simple rescaling of $\mathrm{BR}(\mu\to e\gamma)$. In addition to the off-shell photon contribution, the process receives relevant contributions from the $Z^\prime$ penguin and from box diagrams involving the exotic quark sector. As shown below, these terms significantly affect the expected signal, especially in the region $M_{N_1}\sim M_{W^\prime}$.
The conversion rate, normalized to the muon capture, is
\begin{equation}
\begin{aligned}
\label{eq:mueconversion}
\mathrm{CR}(\mu-e,\mathcal N)
=
\frac{p_e E_e\, m_\mu^3\, G_F^2 \alpha_{\mathrm{em}}^3 Z_{\mathrm{eff}}^4 F_p^2}
{8\pi^2 Z\,\Gamma_{\mathrm{capt}}}
\left|
(Z+N)\,g_{LV}^{(0)} + (Z-N)\,g_{LV}^{(1)}
\right|^2 ,
\end{aligned}
\end{equation}
where $Z$ and $N$ denote the number of protons and neutrons in the target nucleus, $Z_{\mathrm{eff}}$ is the effective atomic charge, $F_p$ is the nuclear form factor, and $\Gamma_{\mathrm{capt}}$ is the muon capture rate. The numerical values of these parameters can be found in~\cite{Arganda:2007jw}.

We define the isoscalar and isovector couplings as
\begin{equation}
\begin{aligned}
g_{LV}^{(0)}
&=
\frac{1}{2}
\sum_{q=u,d,s}
g_{LV(q)}
\left(
G_V^{(q,p)} + G_V^{(q,n)}
\right),
\\
g_{LV}^{(1)}
&=
\frac{1}{2}
\sum_{q=u,d,s}
g_{LV(q)}
\left(
G_V^{(q,p)} - G_V^{(q,n)}
\right),
\end{aligned}
\end{equation}
with the standard quark vector charges
\begin{equation}
\begin{aligned}
&G_V^{(u,p)}=2, \qquad G_V^{(d,p)}=1, \qquad G_V^{(s,p)}=0,
\\
&G_V^{(u,n)}=1, \qquad G_V^{(d,n)}=2, \qquad G_V^{(s,n)}=0.
\end{aligned}
\end{equation}
We decompose the coefficients as
\begin{equation}
\label{eq:muconv_coefs}
\begin{aligned}
g_{LV(q)}
&=
g_{LV(q)}^\gamma
+
g_{LV(q)}^{Z^\prime}
+
g_{LV(q)}^{\mathrm{box}}=\frac{\sqrt2}{G_F}\left(e^2Q_q\,\Delta_\gamma+\Delta_{Z^\prime}^{(q)}+\Delta_{\rm box}^{(u)}\right),
\end{aligned}
\end{equation}
where $q=u,d$\footnote{Heavier quarks are not relevant in pure vector interactions \cite{Abada:2014kba}.}. The photonic contribution $\Delta_\gamma$ contains both dipolar and non-dipolar pieces and depends on the form factors $K_2$ and $K_1$, respectively. $\Delta_{Z^\prime}^{(q)}$ describes the contact interaction between the $Z^\prime$ penguin and the quark current, while $\Delta_{\rm box}^{(u)}$ encodes box contributions involving lepton and quark currents. Explicit expressions for $\Delta_\gamma$, $\Delta_{Z^\prime}$, and $\Delta_{\rm box}^{(u)}$ are given in Appendix~\ref{app:LFV_expressions}.

The last terms in Eq.~(\ref{eq:muconv_coefs}) are absent in the usual dipole approximation and constitute the main sources of deviation from the linear relation with $\mu\to e\gamma$. In this model, the $W^\prime$ boson connects ordinary and exotic fermions, so the relevant quark--lepton box contribution arises from the $u$--$D_1$ transition. By contrast, the gauge symmetry of the model implies that an analogous down-quark contribution is absent.

 Considering only the dipole contribution, we recover the usual linear scaling $\mathrm{CR}(\mu-e, N)\propto \mathrm{BR}(\mu\to e\gamma)$. However, including the non-dipolar photon term and the $Z^\prime$ penguin, sizable deviations emerge, particularly near $M_{N_1}\sim M_{W^\prime}$. The exotic-quark box contribution enhances this effect and may dominate in parts of parameter space. In summary, $\mu$--$e$ conversion in this model is directly sensitive to the exotic quark sector.

\subsection{Results}

\begin{figure}
    \centering
    \includegraphics[width=0.8\linewidth]{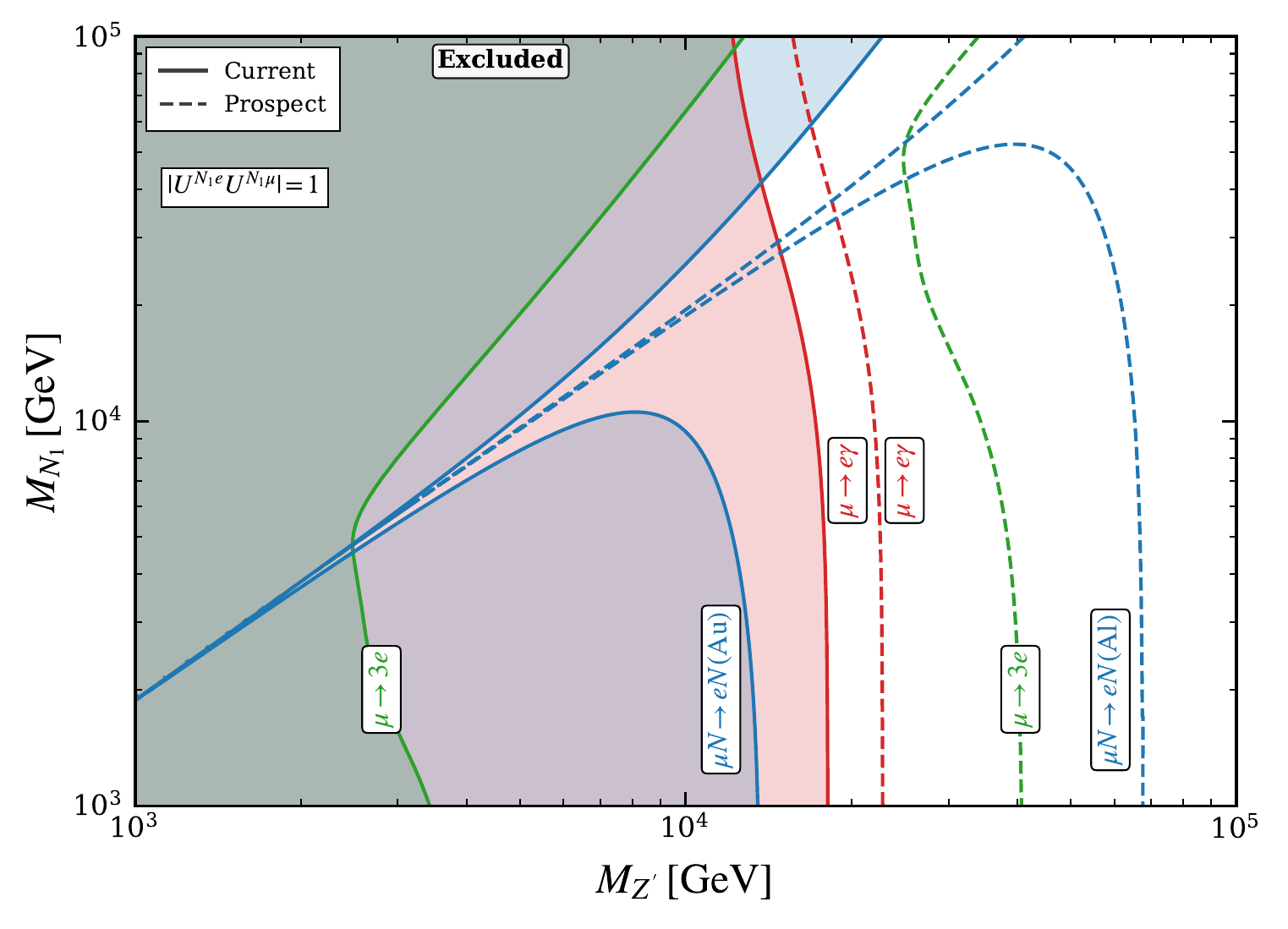}
    \includegraphics[width=0.8\linewidth]{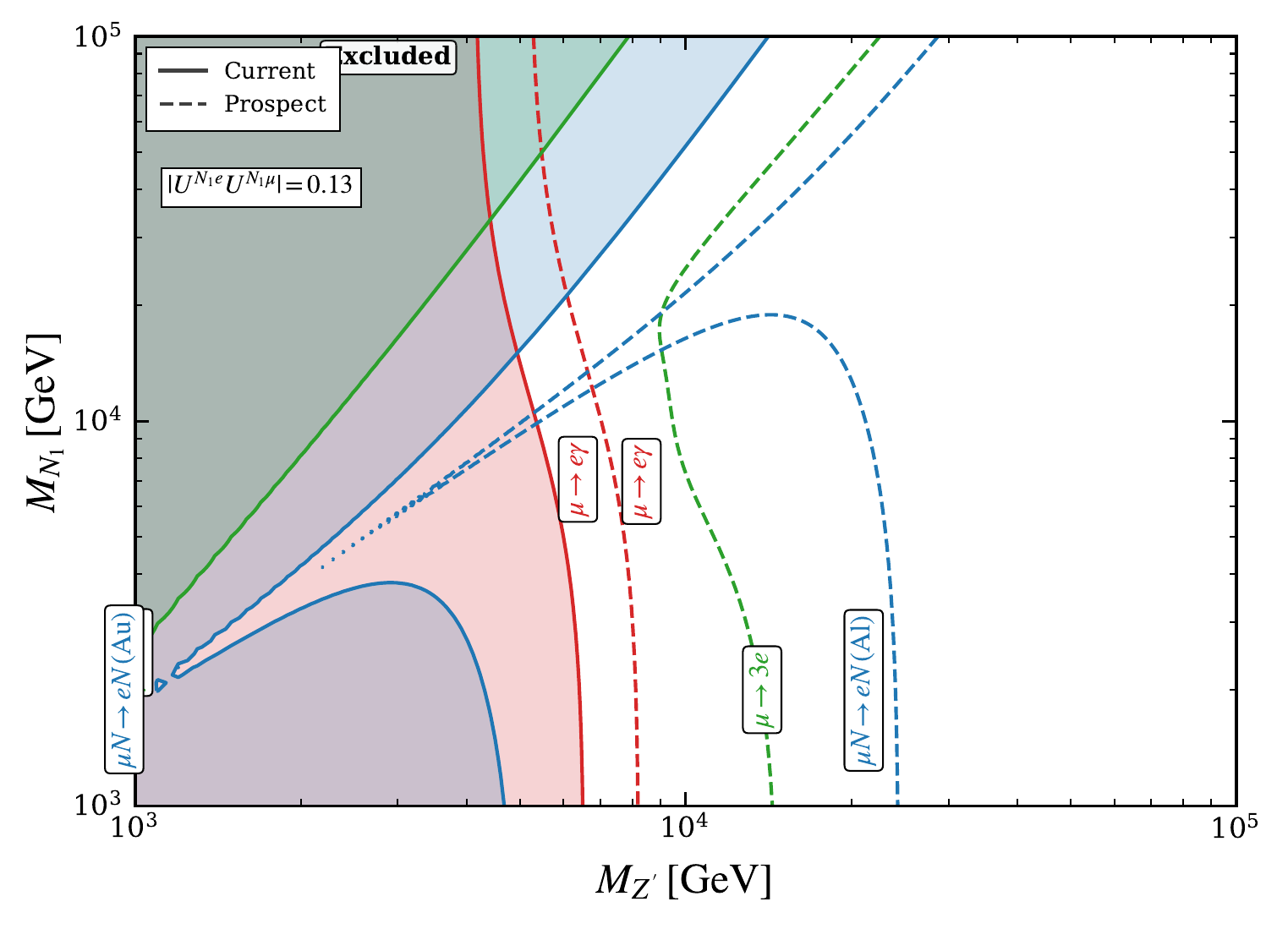}
    \caption{Current and projected sensitivities of the three LFV observables in the $M_{Z'}$--$M_{N_1}$ plane. Solid lines denote the present limits, while dashed lines show the prospects. The upper panel corresponds to the maximum-scope benchmark $|U^{N_1e*}U^{N_1\mu}|=1$, which illustrates the largest LFV reach allowed by the model. The lower panel shows the representative choice $|U^{N_1e*}U^{N_1\mu}|=0.13$, adopted in the global analysis.}
    \label{fig:LFV}
\end{figure}

In this section, we evaluate the three relevant observables
\begin{align}
    \text{BR}(\mu\to e \gamma)=\frac{\Gamma(\mu\to e \gamma)}{\Gamma_{\mu\rm tot}},\quad \text{BR}(\mu\to 3e)=\frac{\Gamma(\mu\to 3e)}{\Gamma_{\mu\rm tot}},\quad \text{CR}(\mu-e,\mathcal N)
\end{align}
using Eqs.~(\ref{eq:mutogaamma}), (\ref{eq>muto3e}), and (\ref{eq:mueconversion}), together with the experimental value $\Gamma_{\mu\mathrm{tot}} = 2.996\times 10^{-19}$ GeV\footnote{Including precise measurement of the muon lifetime.} reported in \cite{ParticleDataGroup:2024cfk}. Once the structure of the model is fixed, the phenomenology of LFV processes is controlled by the masses $M_{Z^\prime}$, $M_{N_1}$, and the lepton--DM mixing $|U^{N_1 e*}U^{N_1 \mu}|$, whereas $M_{W^\prime}$ is not independent because it is tied to the symmetry-breaking scale of the model through Eq.~(\ref{eq:mwmass}). We therefore use $M_{W^\prime}=0.8\,M_{Z^\prime}$. Likewise, $\mu$--$e$ conversion also depends on the heavy-quark mass $M_{D_1}$, which we fix near the symmetry-breaking scale of the model, $M_{D_1}\sim v_\chi$. The aim is to study the current and projected reach of the LFV signal in the $M_{Z^\prime}$--$M_{N_1}$ plane for different lepton--DM mixing.

The results are summarized in \autoref{fig:LFV}, which shows the current and future sensitivity regions for the three LFV observables considered at fixed values of $|U^{N_1 e*}U^{N_1 \mu}|$. Here, the solid lines represent the current experimental upper bounds, whereas the dashed lines correspond to the projected experimental sensitivities. The current limits and prospects are listed in \autoref{tab:signal}. Notably, in the region where the DM candidate is viable, $M_{N_1}<M_{W^\prime}$, all three observables exhibit the linear behavior with respect to $\mu \to e\gamma$, in agreement with previous studies \cite{Arcadi:2017xbo}. In the region where the DM candidate is unstable, $M_{N_1}>M_{W^\prime}$, the contributions from operators beyond the dipole term become more prominent. In particular, the $\mu$--$e$ conversion signal stands out phenomenologically because it provides a direct test of the heavy-quark sector of the model.

The upper panel of \autoref{fig:LFV} shows the benchmark $|U^{N_1 e*}U^{N_1 \mu}|=1$ as a reference for the maximum reach of the model. This scenario is not interpreted as physically preferred, but rather as an illustration of the largest exclusion regions that LFV signals can produce in this model. It shows that the intrinsic flavor mixing reach of the model extends into the tens of TeV regime, thereby highlighting that the model remains experimentally relevant even at mass scales well above the current experimental reach. The $\mu \to e\gamma$ channel is the most restrictive observable in the region where the dark matter candidate is viable, whereas $\mu$--$e$ conversion shows the greatest projected sensitivity and is also the most restrictive observable in the region $M_{N_1}>M_{W^\prime}$, where its exclusion limit even exceeds that from direct detection.

The lower panel of \autoref{fig:LFV} shows the benchmark $|U^{N_1 e*}U^{N_1 \mu}|=0.13$, which is used later in the combined analysis with DM. This scenario, therefore, emerges as one of particular phenomenological interest in view of future experimental progress. In this case, the reduction in leptonic mixing significantly suppresses the LFV signal; however, it still preserves sensitivity in the TeV-region of the mass plane, confirming that LFV observables remain a competitive probe.

Taken together, these results show that the reach of LFV signals depends strongly on both the scale of new physics and the magnitude of exotic leptonic mixing, while also identifying the current and projected hierarchy among the LFV observables. In the next section, we combine these results with the constraints arising from DM and collider phenomenology to determine the globally viable region of the model.

%% file: sections/discussion.tex
\section{Discussion}
\label{discussion}

The phenomenology of the model is determined by the interplay among dark matter, direct detection, LFV, and collider constraints. In this sense, the physically relevant region of parameter space cannot be identified from a single observable, but only through a global analysis of all these sectors.

From the collider perspective, the most important constraint on the heavy gauge sector arises from resonant production of the $Z^\prime$ boson with an integrated luminosity of $139~\mathrm{fb}^{-1}$ \cite{Coutinho:2013lta,Alves:2022hcp,Salazar:2015gxa,Lindner:2016lpp,Arcadi:2017xbo,DeJesus:2020yqx,deJesus:2023lvn,deJesus:2023som}, which restricts the allowed masses of the exotic states that affect the decays of the $Z^\prime$ boson, as well as its mass. The limit adopted in this work, $M_{Z^\prime}\gtrsim 4.374~\mathrm{TeV}$, is consistent with the phenomenological setup considered. By contrast, the charged $W^\prime$ boson does not provide an independent direct bound, because its production at colliders requires the presence of exotic states. 

\begin{figure}
    \centering
    \includegraphics[width=0.8\linewidth]{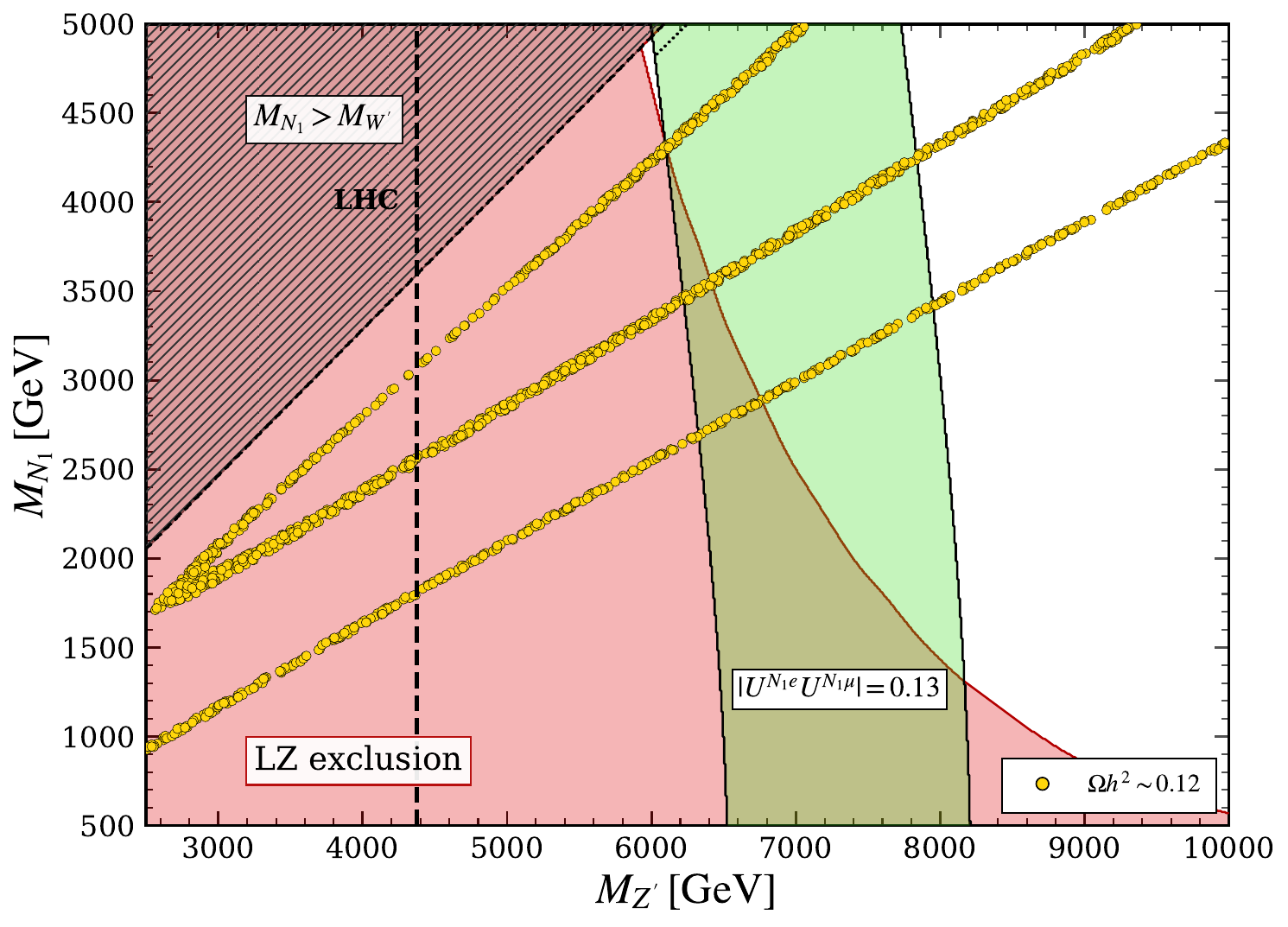}
    \caption{Combined DM and LFV constraints in the $M_{Z'}$--$M_{N_1}$ plane for the benchmark value $|U^{N_1e*}U^{N_1\mu}|=0.13$. The yellow bands reproduce the observed relic-density, the red region is excluded by LZ, in the hatched region with $M_{N_1}>M_{W'}$ the DM candidate is unstable, and the vertical dashed line indicates the collider bound on $M_{Z'}$. The green region shows the parameter space where $\mu\to e\gamma$ reaches its present sensitivity, identifying the most restrictive LFV probe in the current experimental scenario.}
    \label{fig:dmlfv}
\end{figure}
The combination of these constraints is summarized in~\autoref{fig:dmlfv}, where they are projected onto the $(M_{Z^\prime}, M_{N_1})$ plane. Regarding DM, the low mass regime for $N_1$ and $Z'$ is excluded by current LZ limits over DM direct detection \cite{LZ:2024zvo} and is represented by the red region; in addition, the yellow points represent the DM candidate accounting for $100\%$ of the observed relic abundance~\cite{Planck:2018vyg}. To compare DM and LFV phenomenology, we plot in green the area that corresponds to the signal region for $\mu\to{e}\gamma$, i.e., between the current measurement 
\cite{MEGII:2025gzr} and the most promising near term prospect 
\cite{MEGII:2025gzr}. This region is particularly interesting to identify new physics in the near future \cite{COMET:2025sdw}. Notice that the lepton modulates the signal region--DM mixing strength  $\left|U^{N_1 e *} U^{N_1 \mu}\right|$, so an increment of this mixing moves the signal region to mass values of tens of TeV in the $M_{Z'}$--$M_{N_1}$ plane, and make LFV measurements compete even with future multi-ton direct detection experiments, such as DARWIN \cite{Schumann:2015cpa}, which aim to extend the exploration of heavy WIMP scenarios \cite{Billard:2021uyg}.

An important result of this work is that DM viability selects precisely a region of parameter space in which LFV physics remains essentially dipole-dominated. In other words, the region compatible with the relic density, direct detection, and stability of the candidate coincides with the regime in which $\mu \to 3e$ and $\mu$--$e$ conversion remain strongly correlated with $\mu \to e\gamma$, in agreement with previous studies \cite{Arcadi:2017xbo}. However, near the boundary of DM stability, the LFV signal is enhanced, particularly for $\mu$--$e$ conversion events. By contrast, outside the viable DM region, contributions from the $Z^\prime$ penguin and fermionic box diagrams can become decisive, thereby modifying the hierarchy of LFV observables and broadening the phenomenological reach of the model.

In summary, the analysis shows that the model remains highly predictive once all relevant constraints are imposed simultaneously. In the current experimental scenario, $\mu \to e\gamma$ remains the most restrictive LFV signal within the viable region, whereas, in the future, $\mu$--$e$ conversion in nuclei emerges as the most promising channel for probing extended regions of the model.

%% file: sections/conclusion.tex
\section{Conclusion}\label{conclusions}

In this work, we revisit the phenomenology of the 331-LHN model by jointly considering fermionic DM, LFV, and collider constraints. In particular, we update the constraints imposed by direct detection experiments searching for DM and by collider searches for new signals, which push the DM candidate mass above roughly $10$ TeV. The LFV analysis of the model is extended through a more complete treatment of the processes $\mu \to e\gamma$, $\mu \to 3e$, and $\mu$--$e$ conversion, explicitly incorporating the contributions from the $Z^\prime$ penguin and fermionic box diagrams.

One of the main conceptual results of this article is that, although LFV processes receive additional contributions beyond the dipole approximation, the region of parameter space compatible with DM viability coincides precisely with the regime in which LFV physics remains essentially dipole-dominated. In this sense, previous studies accurately captured the model's behavior in the most physically motivated region. The novelty of the present study is to show that, outside this scenario, contributions from the exotic-particle sector, particularly from the exotic quark sector, can substantially modify the hierarchy among observables and broaden the phenomenological reach of the model.

From a phenomenological perspective, the combination of experimental constraints on new physics selects a relatively narrow region of parameter space, thereby making the model highly predictive. Within this region, $\mu \to e\gamma$ emerges as the most restrictive LFV signal in the current experimental scenario and in the near-term experimental outlook. Finally, $\mu$--$e$ conversion in nuclei appears as the most promising channel in the long term, both because of its projected experimental sensitivity and because of its dependence on the heavy-quark sector. Taken together, these results show that the 331-LHN model remains a viable, predictive, and experimentally accessible framework for exploring the connection between dark matter and lepton flavor violation.

%% file: sections/acknowledgments.tex
\section*{Acknowledgments}

This work is supported by Simons Foundation (Award Number:1023171-RC), FAPESP Grants 2021/01089-1, 2023/01197-4, ICTP-SAIFR FAPESP Grants 2021/14335-0, 
CNPq grants 403521/2024-6, 408295/2021-0, 403521/2024-6, 406919/2025-9, 351851/2025-9, ANID-Millennium Science Initiative Program ICN2019\_044, and IIF-FINEP grant 213/2024. P. E. is supported by CNPq grant No. 151612/2024-2. 
J. E. was supported by ANID (Chile) under FONDECYT Grant No. 3260491. J. E. also acknowledges Universidad Técnica Federico Santa María (UTFSM) for support during part of this work.

%% file: sections/appendix.tex
\section{Expressions for LFV processes}
\label{app:LFV_expressions}

In this section, we detail the ingredients for calculating processes that violate lepton flavor in this model, based on the compilation of contributions presented in \cite{Abada:2014kba}. As detailed in the reference, we work in the Feynman gauge. Therefore, the total contribution must include both the exotic gauge bosons and their Goldstone bosons. We further assume that the lightest heavy neutral fermion, $N_1$, dominates LFV. For simplicity, we define
\begin{align}
\Lambda\equiv g_{N_1e}^* g_{N_1\mu}
=\frac{g^2}{8}U^{N_1 e *} U^{N_1 \mu}.
\end{align}

The interaction terms in the following Lagrangian render the trilineal coupling among $W^\prime$, $Z^\prime$, $\gamma$, and the Goldstones $G^{\prime 0}, G^{\prime \pm}$
\begin{align}
    \mathcal{L} = G^{\prime +}\,\bar N \Big(\Gamma^{L}_{N\ell G^{\prime +}}P_L+\Gamma^{R}_{N\ell G^{\prime +}}P_R\Big)\ell+G^{\prime -}\,\bar\ell\Big(\Gamma^{L}_{\ell N G^{\prime -}}P_L+\Gamma^{R}_{\ell N G^{\prime -}}P_R\Big)N,
\end{align}
with
\begin{align}
    \Gamma^{L}_{\ell_\alpha N_i G^{\prime -}}
    &=\big(\Gamma^{R}_{N_i\ell_\alpha G^{\prime +}}\big)^{*}
    =-\frac{g}{2\sqrt{2}}\frac{m_{\ell_\alpha}}{M_{W^\prime}}\,U^{N_i\ell *},
    \label{eq:Gprime-conj-L}\\
    \Gamma^{R}_{\ell_\alpha N_i G^{\prime -}}
    &=\big(\Gamma^{L}_{N_i\ell_\alpha G^{\prime +}}\big)^{*}
    =+\frac{g}{2\sqrt{2}}\frac{M_{N_i}}{M_{W^\prime}}\,U^{N\ell_\alpha *}.
\end{align}
On the other hand, the photonic vertices are
\begin{align}
    \Gamma_{G'G'\gamma}=-e,\quad\Gamma_{W'W'\gamma}=e,\quad\Gamma_{G'\gamma W'}=eM_{W^\prime}.
\end{align}

\subsection{Photon contribution}

The effective photonic LFV Lagrangian is
\begin{align}
\mathcal L_{\ell\ell\gamma}=
 e\,\bar\ell_\beta
\left[
\gamma^\mu\left(K_1^L P_L+K_1^R P_R\right)
+i m_{\ell_\alpha}\sigma^{\mu\nu}q_\nu\left(K_2^L P_L+K_2^R P_R\right)
\right]
\ell_\alpha A_\mu+\text{h.c.}
\end{align}
We show the diagrams that contribute to LFV in \autoref{fig:mu_e_gamma}. At leading order, $m_e,m_\mu\ll m_{N_1},m_{W^\prime}$, and in the approximation dominated by the gauge sector, the non-dipole and dipole terms, $K_1$ and $K_2$, respectively, are
\begin{align}
&K_1^L=\frac{\Lambda}{(4\pi)^2M_{W'}^2}F_1(x),\quad K_1^R\simeq 0,\\
&K_2^R=\frac{\Lambda}{(4\pi)^2M_{W'}^2}F_2(x),\quad K_2^L\simeq 0,
\end{align}
where the non-dipole and dipole loop functions, $F_1(x)$ and $F_2(x)$, respectively, are
\begin{align}&F_1(x)=\frac{x\left(11x^3-18x^2+9x-2-6x^3\ln x\right)}{36(x-1)^4}\\
&F_2(x)=\frac{-10+43x-78x^2+49x^3-4x^4-18x^3\ln{(x)}}{12(x-1)^4},
\end{align}
with $x\equiv \frac{M_{N_1}^2}{M_{W^\prime}^2}$.
The combination that enters the photon-induced vector operator is
\begin{align}
\Delta_\gamma\equiv K_1^L-K_2^R,
\end{align}
Therefore, the photon contribution to the four-fermion interaction is
\begin{align}
 &g_{LV(q)}^{(\gamma)}=\frac{\sqrt2}{G_F}e^2Q_q\,\Delta_\gamma,\quad g_{RV(q)}^{(\gamma)}\simeq 0,
 \quad
 g_{LS(q)}^{(\gamma)}=g_{RS(q)}^{(\gamma)}=0.
\end{align}

\subsection{Contribution of the $Z^\prime$}

The neutral chiral interactions between the $Z^\prime$ and fermions are given by
\begin{align}
 g_L^{\prime f}=g_V^{\prime f}-g_A^{\prime f},
 \qquad
 g_R^{\prime f}=g_V^{\prime f}+g_A^{\prime f}.
\end{align}
From Eq.~(\ref{eq:fermion_interactions}), we obtain the couplings
\begin{align}
 g_L^{\prime \ell}=\frac{2s_W^2-1}{\sqrt{3-4s_W^2}},
 \qquad
 g_R^{\prime \ell}=\frac{2s_W^2}{\sqrt{3-4s_W^2}},
\end{align}
\begin{align}
 g_L^{\prime N}=\frac{8}{9}\sqrt{3-4s_W^2},
 \qquad
 g_R^{\prime N}=0,
\end{align}
\begin{align}
 g_V^{\prime u}=\frac{3-8s_W^2}{6\sqrt{3-4s_W^2}},
 \qquad
 g_V^{\prime d}=g_V^{\prime s}=\frac{3-2s_W^2}{6\sqrt{3-4s_W^2}}.
\end{align}
On the other hand, the relevant trilinear couplings are \cite{Long:1995ctv}
\begin{equation}
    \begin{aligned}
    &g_{ZWW}=\sqrt{3-4s_W^2},
     \qquad
    g_{ZWG}=\sqrt{3-4s_W^2},\\ 
    &g_{ZGG}=\frac{1-2s_W^2}{\sqrt{3-4s_W^2}}\left[1+\left(\frac{gv}{2M_{W^\prime}}\right)^2\right].
    \end{aligned}
\end{equation}
We parametrize the effective vertex as
\begin{align}
\mathcal L_{\ell\ell Z'}
=-\frac{g}{2c_W}Z'_\mu\,\bar \ell\gamma^\mu\left(V_L^{Z'}P_L+V_R^{Z'}P_R\right)\ell+\text{h.c.}
\end{align}
Again, at leading order, the vertex operator $V^{Z^\prime}$ takes the form
\begin{align}
V_L^{Z'}=\frac{\Lambda}{(4\pi)^2}F^\prime(x),\quad V_R^{Z'}\simeq 0,
\end{align}
where the loop function $F^\prime(x)$ is
\begin{equation}
\label{eq:zloopfunction}
    \begin{aligned}
    F^\prime(x)=
    &\left[
    \frac{2s_W^2-1}{\sqrt{3-4s_W^2}}
    +\frac{1-2s_W^2}{\sqrt{3-4s_W^2}}\left[1+\left(\frac{gv}{2M_{W^\prime}}\right)^2\right]
    \right]
    \frac{x\left(2x^2\ln x-3x^2+4x-1\right)}{4(x-1)^2}
    \\
    &+\frac{8}{9}\sqrt{3-4s_W^2}
    \left[
    \frac{x^2\left(x-\ln x-1\right)}{(x-1)^2}
    -\frac{2x^2\ln x-x^2-4x\ln x+4x-3}{2(x-1)^2}
    \right]\\
    &+\frac{2s_W^2-1}{\sqrt{3-4s_W^2}}
    \frac{2x^2\ln x-x^2+1}{2(x-1)^2}
    -\sqrt{3-4s_W^2}
    \frac{2x\left(x\ln x-x+1\right)}{(x-1)^2}
    \\
    &+\sqrt{3-4s_W^2}
    \frac{3\left(2x^2\ln x-3x^2+4x-1\right)}{2(x-1)^2},
    \end{aligned}
\end{equation}
with $x\equiv \frac{M_{N_1}^2}{M_{W^\prime}^2}$.
Finally, the contribution to the four-fermion operator is
\begin{align}
{\Delta_{Z^\prime}^{(f)}}_{(L/R)}=\frac{g^2}{4c_W^2M_{Z'}^2}V_L^{Z'}g_{(L/R)}^{\prime f}.
\end{align}
So, for $f=\ell,q$, as in Figs \ref{fig:mu_3e_diagrams}(b) and \ref{fig:mu_e_conv_diagrams}(b)

\begin{align}
{\Delta_{Z^\prime}^{(\ell)}}_{(L/R)}=\frac{g^2}{4c_W^2M_{Z'}^2}V_L^{Z'}g_{(L/R)}^{\prime \ell},\quad
{\Delta_{Z^\prime}^{(q)}}=\frac{g^2}{4c_W^2M_{Z'}^2}V_L^{Z'}g_V^{\prime q}.
\end{align}

\subsection{Box contributions}

In a 331 model, the $W^\prime$ boson always interacts with one ordinary fermion and one exotic fermion. For charged leptons, the box is mediated by heavy neutrinos as can be seen in Fig.~\ref{fig:mu_3e_diagrams}(c), whereas for light quarks, this implies the existence of a direct box with a $u$ quark through $u-D_1$, as shown in Fig.~\ref{fig:mu_e_conv_diagrams}(c), whereas no analogous box appears for $d,s$ without introducing additional mixing between ordinary and exotic quarks.

Therefore, the contribution of the lepton--lepton and quark--lepton box diagrams to the four-fermion interaction is

\begin{align}
&\Delta^{(f)}_{\rm box}=\left|\Gamma_{W^\prime F f}\right|^2\frac{\Lambda}{(4\pi)^2M_{W'}^2}B(x,y)\quad (\rm{only\,left})\\
&\Delta^{(f)}_{\rm box,R}\simeq 0,
\end{align}
where $\Gamma_{W^\prime F f}=\frac{g}{2\sqrt{2}}$ is the trilinear interaction between $W^\prime$, the ordinary fermion $f$ and the exotic fermion $F$ and the loop function $B(x,y)$ is given by
\begin{align}
B(x,y)=(16+xy)\widehat D_{27}(x,y)+2xy\widehat D_0(x,y),
\end{align}
with
\begin{align}
&\widehat D_0(x,y)
=
-\frac{x\,\ln x}{(x-1)^2(x-y)}
+\frac{y\,\ln y}{(x-y)(y-1)^2}
-\frac{1}{(x-1)(y-1)},\\
&\widehat D_{27}(x,y)
=
-\frac{x^2\,\ln x}{4(x-1)^2(x-y)}
+\frac{y^2\,\ln y}{4(x-y)(y-1)^2}
-\frac{1}{4(x-1)(y-1)},
\end{align}
here $x\equiv \frac{M_{F_1}^2}{M_{W^\prime}^2}$ and $y\equiv \frac{M_{F_2}^2}{M_{W^\prime}^2}$. In the case $M_{F_1}=M_{F_2}$ the loop function simplifies to

\begin{equation}
\begin{aligned}
\widehat D_0(x,x)
=
\frac{(x^2-1)\ln x-2x^2+4x-2}{(x-1)^4},
\end{aligned}
\end{equation}
\begin{equation}
\begin{aligned}
\widehat D_{27}(x,x)
=
\frac{1-x^2+2x\ln x}{4(x-1)^3}.
\end{aligned}
\end{equation}

%% file: references.bib
@article{Arcadi:2024ukq,
    author = "Arcadi, Giorgio and Cabo-Almeida, David and Dutra, Ma{\'\i}ra and Ghosh, Pradipta and Lindner, Manfred and Mambrini, Yann and Neto, Jacinto P. and Pierre, Mathias and Profumo, Stefano and Queiroz, Farinaldo S.",
    title = "{The Waning of the WIMP: Endgame?}",
    eprint = "2403.15860",
    archivePrefix = "arXiv",
    primaryClass = "hep-ph",
    doi = "10.1140/epjc/s10052-024-13672-y",
    journal = "Eur. Phys. J. C",
    volume = "85",
    number = "2",
    pages = "152",
    year = "2025"
}

@article{Bertone:2016nfn,
    author = "Bertone, Gianfranco and Hooper, Dan",
    title = "{History of dark matter}",
    eprint = "1605.04909",
    archivePrefix = "arXiv",
    primaryClass = "astro-ph.CO",
    reportNumber = "FERMILAB-PUB-16-157-A",
    doi = "10.1103/RevModPhys.90.045002",
    journal = "Rev. Mod. Phys.",
    volume = "90",
    number = "4",
    pages = "045002",
    year = "2018"
}

@article{deSwart:2017heh,
    author = "de Swart, Jaco and Bertone, Gianfranco and van Dongen, Jeroen",
    title = "{How Dark Matter Came to Matter}",
    eprint = "1703.00013",
    archivePrefix = "arXiv",
    primaryClass = "astro-ph.CO",
    doi = "10.1038/s41550017-0059",
    journal = "Nature Astron.",
    volume = "1",
    pages = "0059",
    year = "2017"
}

@article{Goodman:1984dc,
    author = "Goodman, Mark W. and Witten, Edward",
    editor = "Srednicki, M. A.",
    title = "{Detectability of Certain Dark Matter Candidates}",
    reportNumber = "Print-85-0030 (PRINCETON)",
    doi = "10.1103/PhysRevD.31.3059",
    journal = "Phys. Rev. D",
    volume = "31",
    pages = "3059",
    year = "1985"
}

@article{Feruglio:2015jfa,
    author = "Feruglio, Ferruccio",
    title = "{Pieces of the Flavour Puzzle}",
    eprint = "1503.04071",
    archivePrefix = "arXiv",
    primaryClass = "hep-ph",
    doi = "10.1140/epjc/s10052-015-3576-5",
    journal = "Eur. Phys. J. C",
    volume = "75",
    number = "8",
    pages = "373",
    year = "2015"
}

@article{Novichkov:2021evw,
    author = "Novichkov, P. P. and Penedo, J. T. and Petcov, S. T.",
    title = "{Fermion mass hierarchies, large lepton mixing and residual modular symmetries}",
    eprint = "2102.07488",
    archivePrefix = "arXiv",
    primaryClass = "hep-ph",
    reportNumber = "SISSA 09/2021/FISI, IPMU21-0009, CFTP/21-003",
    doi = "10.1007/JHEP04(2021)206",
    journal = "JHEP",
    volume = "04",
    pages = "206",
    year = "2021"
}

@article{Pisano:1996ht,
    author = "Pisano, Felice",
    title = "{A Simple solution for the flavor question}",
    eprint = "hep-ph/9609358",
    archivePrefix = "arXiv",
    reportNumber = "IFT-P-031-96",
    doi = "10.1142/S0217732396002630",
    journal = "Mod. Phys. Lett. A",
    volume = "11",
    pages = "2639--2647",
    year = "1996"
}

@article{Dias:2004dc,
    author = "Dias, Alex G. and Martinez, R. and Pleitez, V.",
    title = "{Concerning the Landau pole in 3-3-1 models}",
    eprint = "hep-ph/0407141",
    archivePrefix = "arXiv",
    doi = "10.1140/epjc/s2004-02083-0",
    journal = "Eur. Phys. J. C",
    volume = "39",
    pages = "101--107",
    year = "2005"
}

@article{Dias:2004wk,
    author = "Dias, Alex Gomes",
    title = "{Evading the few TeV perturbative limit in 3-3-1 models}",
    eprint = "hep-ph/0412163",
    archivePrefix = "arXiv",
    doi = "10.1103/PhysRevD.71.015009",
    journal = "Phys. Rev. D",
    volume = "71",
    pages = "015009",
    year = "2005"
}

@article{deSalas:2020pgw,
    author = "de Salas, P. F. and Forero, D. V. and Gariazzo, S. and Mart{\'\i}nez-Mirav{\'e}, P. and Mena, O. and Ternes, C. A. and T{\'o}rtola, M. and Valle, J. W. F.",
    title = "{2020 global reassessment of the neutrino oscillation picture}",
    eprint = "2006.11237",
    archivePrefix = "arXiv",
    primaryClass = "hep-ph",
    doi = "10.1007/JHEP02(2021)071",
    journal = "JHEP",
    volume = "02",
    pages = "071",
    year = "2021"
}

@article{Capozzi:2025wyn,
    author = "Capozzi, Francesco and Giar{\`e}, William and Lisi, Eligio and Marrone, Antonio and Melchiorri, Alessandro and Palazzo, Antonio",
    title = "{Neutrino masses and mixing: Entering the era of subpercent precision}",
    eprint = "2503.07752",
    archivePrefix = "arXiv",
    primaryClass = "hep-ph",
    doi = "10.1103/PhysRevD.111.093006",
    journal = "Phys. Rev. D",
    volume = "111",
    number = "9",
    pages = "093006",
    year = "2025"
}

@article{Esteban:2024eli,
    author = "Esteban, Ivan and Gonzalez-Garcia, M. C. and Maltoni, Michele and Martinez-Soler, Ivan and Pinheiro, Jo{\~a}o Paulo and Schwetz, Thomas",
    title = "{NuFit-6.0: updated global analysis of three-flavor neutrino oscillations}",
    eprint = "2410.05380",
    archivePrefix = "arXiv",
    primaryClass = "hep-ph",
    reportNumber = "IFT-UAM/CSIC-24-140, YITP-SB-2024-24, IPPP/24/64, IPPP/24/64, IFT-UAM/CSIC-24-140, YITP-SB-2024-24",
    doi = "10.1007/JHEP12(2024)216",
    journal = "JHEP",
    volume = "12",
    pages = "216",
    year = "2024"
}

@article{ParticleDataGroup:2024cfk,
    author = "Navas, S. and others",
    collaboration = "Particle Data Group",
    title = "{Review of particle physics}",
    doi = "10.1103/PhysRevD.110.030001",
    journal = "Phys. Rev. D",
    volume = "110",
    number = "3",
    pages = "030001",
    year = "2024"
}

@article{Planck:2018vyg,
    author = "Aghanim, N. and others",
    collaboration = "Planck",
    title = "{Planck 2018 results. VI. Cosmological parameters}",
    eprint = "1807.06209",
    archivePrefix = "arXiv",
    primaryClass = "astro-ph.CO",
    doi = "10.1051/0004-6361/201833910",
    journal = "Astron. Astrophys.",
    volume = "641",
    pages = "A6",
    year = "2020",
    note = "[Erratum: Astron.Astrophys. 652, C4 (2021)]"
}

@article{LZ:2024zvo,
    author = "Aalbers, J. and others",
    collaboration = "LZ",
    title = "{Dark Matter Search Results from 4.2{\,}{\,}Tonne-Years of Exposure of the LUX-ZEPLIN (LZ) Experiment}",
    eprint = "2410.17036",
    archivePrefix = "arXiv",
    primaryClass = "hep-ex",
    reportNumber = "FERMILAB-PUB-24-0796-V",
    doi = "10.1103/4dyc-z8zf",
    journal = "Phys. Rev. Lett.",
    volume = "135",
    number = "1",
    pages = "011802",
    year = "2025"
}

@article{Mizukoshi:2010ky,
    author = "Mizukoshi, J. K. and de S. Pires, C. A. and Queiroz, F. S. and Rodrigues da Silva, P. S.",
    title = "{WIMPs in a 3-3-1 model with heavy Sterile neutrinos}",
    eprint = "1010.4097",
    archivePrefix = "arXiv",
    primaryClass = "hep-ph",
    doi = "10.1103/PhysRevD.83.065024",
    journal = "Phys. Rev. D",
    volume = "83",
    pages = "065024",
    year = "2011"
}

@article{Arcadi:2017xbo,
    author = "Arcadi, Giorgio and Ferreira, C. P. and Goertz, Florian and Guzzo, M. M. and Queiroz, Farinaldo S. and Santos, A. C. O.",
    title = "{Lepton Flavor Violation Induced by Dark Matter}",
    eprint = "1712.02373",
    archivePrefix = "arXiv",
    primaryClass = "hep-ph",
    doi = "10.1103/PhysRevD.97.075022",
    journal = "Phys. Rev. D",
    volume = "97",
    number = "7",
    pages = "075022",
    year = "2018"
}

@article{Salazar:2015gxa,
    author = "Salazar, Camilo and Benavides, Richard H. and Ponce, William A. and Rojas, Eduardo",
    title = "{LHC Constraints on 3-3-1 Models}",
    eprint = "1503.03519",
    archivePrefix = "arXiv",
    primaryClass = "hep-ph",
    doi = "10.1007/JHEP07(2015)096",
    journal = "JHEP",
    volume = "07",
    pages = "096",
    year = "2015"
}

@article{Lindner:2016lpp,
    author = "Lindner, Manfred and Queiroz, Farinaldo S. and Rodejohann, Werner",
    title = "{Dilepton bounds on left{\textendash}right symmetry at the LHC run II and neutrinoless double beta decay}",
    eprint = "1604.07419",
    archivePrefix = "arXiv",
    primaryClass = "hep-ph",
    doi = "10.1016/j.physletb.2016.08.068",
    journal = "Phys. Lett. B",
    volume = "762",
    pages = "190--195",
    year = "2016"
}

@article{DeJesus:2020yqx,
    author = "De Jesus, A. S. and Kovalenko, S. and Queiroz, F. S. and Siqueira, C. and Sinha, K.",
    title = "{Vectorlike leptons and inert scalar triplet: Lepton flavor violation, $g-2$, and collider searches}",
    eprint = "2004.01200",
    archivePrefix = "arXiv",
    primaryClass = "hep-ph",
    doi = "10.1103/PhysRevD.102.035004",
    journal = "Phys. Rev. D",
    volume = "102",
    number = "3",
    pages = "035004",
    year = "2020"
}

@article{Alves:2022hcp,
    author = "Alves, A. and Duarte, L. and Kovalenko, S. and Oviedo-Torres, Y. M. and Queiroz, F. S. and Villamizar, Y. S.",
    title = "{Constraining 3-3-1 models at the LHC and future hadron colliders}",
    eprint = "2203.02520",
    archivePrefix = "arXiv",
    primaryClass = "hep-ph",
    doi = "10.1103/PhysRevD.106.055027",
    journal = "Phys. Rev. D",
    volume = "106",
    number = "5",
    pages = "055027",
    year = "2022"
}

@article{deJesus:2023lvn,
    author = "de Jesus, A. S. and Kovalenko, S. and de Melo, T. B. and Neto, J. P. and Oviedo-Torres, Y. M. and Queiroz, F. S. and Villamizar, Y. S. and Zerwekh, A. R.",
    title = "{On the role of LHC and HL-LHC in constraining flavor changing neutral currents}",
    eprint = "2304.00041",
    archivePrefix = "arXiv",
    primaryClass = "hep-ph",
    doi = "10.1016/j.physletb.2023.138419",
    journal = "Phys. Lett. B",
    volume = "849",
    pages = "138419",
    year = "2024"
}

@article{deJesus:2023som,
    author = "de Jesus, {\'A}. S. and Queiroz, F. S. and Valle, J. W. F. and Villamizar, Y.",
    title = "{Vector-Like Fermions and Inert Scalar Solutions to the Muon g-2 Anomaly and collider probes at the HL-LHC and FCC-hh}",
    eprint = "2312.03851",
    archivePrefix = "arXiv",
    primaryClass = "hep-ph",
    month = "12",
    year = "2023"
}

@article{Coutinho:2013lta,
    author = "Coutinho, Y. A. and Salustino Guimar{\~a}es, V. and Nepomuceno, A. A.",
    title = "{Bounds on Z' from 3-3-1 model at the LHC energies}",
    eprint = "1304.7907",
    archivePrefix = "arXiv",
    primaryClass = "hep-ph",
    doi = "10.1103/PhysRevD.87.115014",
    journal = "Phys. Rev. D",
    volume = "87",
    number = "11",
    pages = "115014",
    year = "2013"
}

@article{MEGII:2025gzr,
    author = "Afanaciev, K. and others",
    collaboration = "MEG II",
    title = "{New limit on the ${\mu ^+ \rightarrow e^+ \gamma }$ decay with the MEG II experiment}",
    eprint = "2504.15711",
    archivePrefix = "arXiv",
    primaryClass = "hep-ex",
    doi = "10.1140/epjc/s10052-025-14906-3",
    journal = "Eur. Phys. J. C",
    volume = "85",
    number = "10",
    pages = "1177",
    year = "2025",
    note = "[Erratum: Eur.Phys.J.C 85, 1317 (2025)]"
}

@article{SINDRUM:1987nra,
    author = "Bellgardt, U. and others",
    collaboration = "SINDRUM",
    title = "{Search for the Decay $\mu^+ \to e^+ e^+ e^-$}",
    reportNumber = "SIN-PR-87-09",
    doi = "10.1016/0550-3213(88)90462-2",
    journal = "Nucl. Phys. B",
    volume = "299",
    pages = "1--6",
    year = "1988"
}

@article{COMET:2025sdw,
    author = "Aoki, M. and others",
    collaboration = "COMET, MEG, Mu2e, Mu3e",
    title = "{Charged Lepton Flavour Violations searches with muons: present and future}",
    eprint = "2503.22461",
    archivePrefix = "arXiv",
    primaryClass = "hep-ex",
    reportNumber = "FERMILAB-PUB-25-0213-PPD",
    month = "3",
    year = "2025"
}

@article{COMET:2018auw,
    author = "Abramishvili, R. and others",
    collaboration = "COMET",
    title = "{COMET Phase-I Technical Design Report}",
    eprint = "1812.09018",
    archivePrefix = "arXiv",
    primaryClass = "physics.ins-det",
    doi = "10.1093/ptep/ptz125",
    journal = "PTEP",
    volume = "2020",
    number = "3",
    pages = "033C01",
    year = "2020"
}

@article{Abada:2014kba,
    author = "Abada, A. and Krauss, Manuel E. and Porod, W. and Staub, F. and Vicente, A. and Weiland, Cedric",
    title = "{Lepton flavor violation in low-scale seesaw models: SUSY and non-SUSY contributions}",
    eprint = "1408.0138",
    archivePrefix = "arXiv",
    primaryClass = "hep-ph",
    reportNumber = "LPT-ORSAY-14-43, BONN-TH-14-11, IFT-UAM-CSIC-14-061, FTUAM-14-25",
    doi = "10.1007/JHEP11(2014)048",
    journal = "JHEP",
    volume = "11",
    pages = "048",
    year = "2014"
}

@article{Long:1995ctv,
    author = "Long, Hoang Ngoc",
    title = "{The 331 model with right handed neutrinos}",
    eprint = "hep-ph/9504274",
    archivePrefix = "arXiv",
    reportNumber = "ITP-95-10-HANOI, IC-95-118, IC/95/118",
    doi = "10.1103/PhysRevD.53.437",
    journal = "Phys. Rev. D",
    volume = "53",
    pages = "437--445",
    year = "1996"
}

@article{Long:1996rfd,
    author = "Long, Hoang Ngoc",
    title = "{SU(3)-L x U(1)-N model for right-handed neutrino neutral currents}",
    eprint = "hep-ph/9607439",
    archivePrefix = "arXiv",
    doi = "10.1103/PhysRevD.54.4691",
    journal = "Phys. Rev. D",
    volume = "54",
    pages = "4691--4693",
    year = "1996"
}

@ARTICLE{1980ApJ...238..471R,
       author = {{Rubin}, V.~C. and {Ford}, W.~K., Jr. and {Thonnard}, N.},
        title = "{Rotational properties of 21 SC galaxies with a large range of luminosities and radii, from NGC 4605 (R=4kpc) to UGC 2885 (R=122kpc).}",
      journal = {apj},
         year = 1980,
        month = jun,
       volume = {238},
        pages = {471-487},
          doi = {10.1086/158003},
       adsurl = {https://ui.adsabs.harvard.edu/abs/1980ApJ...238..471R}
}

@article{HESS:2022ygk,
    author = "Abdalla, H. and others",
    collaboration = "H.E.S.S.",
    title = "{Search for Dark Matter Annihilation Signals in the H.E.S.S. Inner Galaxy Survey}",
    eprint = "2207.10471",
    archivePrefix = "arXiv",
    primaryClass = "astro-ph.HE",
    doi = "10.1103/PhysRevLett.129.111101",
    journal = "Phys. Rev. Lett.",
    volume = "129",
    number = "11",
    pages = "111101",
    year = "2022"
}

@article{Angel:2025xwb,
    author = "Angel, Lucia and Borah, Debasish and Neto, Jacinto P. and Queiroz, Farinaldo S. and de Souza, Vitor",
    title = "{Constraining Effective Field Theories for dark matter candidates annihilating into gamma-ray lines with CTAO}",
    eprint = "2509.08050",
    archivePrefix = "arXiv",
    primaryClass = "hep-ph",
    month = "9",
    year = "2025"
}

@article{Angel:2023rdd,
    author = "Angel, Lucia and Gambini, Guillermo and Guedes, Leticia and Queiroz, Farinaldo S. and de Souza, Vitor",
    title = "{Constraining gamma-ray lines from dark matter annihilation using Fermi-LAT and H.E.S.S.~data}",
    eprint = "2311.17827",
    archivePrefix = "arXiv",
    primaryClass = "hep-ph",
    doi = "10.1088/1475-7516/2024/04/028",
    journal = "JCAP",
    volume = "04",
    pages = "028",
    year = "2024"
}

@article{CMS:2024zqs,
    author = "Hayrapetyan, Aram and others",
    collaboration = "CMS",
    title = "{Dark sector searches with the CMS experiment}",
    eprint = "2405.13778",
    archivePrefix = "arXiv",
    primaryClass = "hep-ex",
    reportNumber = "CMS-EXO-23-005, CERN-EP-2024-106",
    doi = "10.1016/j.physrep.2024.09.013",
    journal = "Phys. Rept.",
    volume = "1115",
    pages = "448--569",
    year = "2025"
}

@article{Giagu:2019fmp,
    author = "Giagu, Stefano",
    title = "{WIMP Dark Matter Searches With the ATLAS Detector at the LHC}",
    doi = "10.3389/fphy.2019.00075",
    journal = "Front. in Phys.",
    volume = "7",
    pages = "75",
    year = "2019"
}

@article{XENON:2025vwd,
    author = "Aprile, E. and others",
    collaboration = "XENON",
    title = "{WIMP Dark Matter Search Using a 3.1 Tonne-Year Exposure of the XENONnT Experiment}",
    eprint = "2502.18005",
    archivePrefix = "arXiv",
    primaryClass = "hep-ex",
    doi = "10.1103/msw4-t342",
    journal = "Phys. Rev. Lett.",
    volume = "135",
    number = "22",
    pages = "221003",
    year = "2025"
}

@article{PandaX:2024qfu,
    author = "Bo, Zihao and others",
    collaboration = "PandaX",
    title = "{Dark Matter Search Results from 1.54{\,}{\,}Tonne{\textperiodcentered}Year Exposure of PandaX-4T}",
    eprint = "2408.00664",
    archivePrefix = "arXiv",
    primaryClass = "hep-ex",
    doi = "10.1103/PhysRevLett.134.011805",
    journal = "Phys. Rev. Lett.",
    volume = "134",
    number = "1",
    pages = "011805",
    year = "2025"
}

@article{XLZD:2024nsu,
    author = "Aalbers, J. and others",
    collaboration = "XLZD",
    title = "{The XLZD Design Book: towards the next-generation liquid xenon observatory for dark matter and neutrino physics}",
    eprint = "2410.17137",
    archivePrefix = "arXiv",
    primaryClass = "hep-ex",
    doi = "10.1140/epjc/s10052-025-14810-w",
    journal = "Eur. Phys. J. C",
    volume = "85",
    number = "10",
    pages = "1192",
    year = "2025"
}

@article{DARWIN:2016hyl,
    author = "Aalbers, J. and others",
    collaboration = "DARWIN",
    title = "{DARWIN: towards the ultimate dark matter detector}",
    eprint = "1606.07001",
    archivePrefix = "arXiv",
    primaryClass = "astro-ph.IM",
    doi = "10.1088/1475-7516/2016/11/017",
    journal = "JCAP",
    volume = "11",
    pages = "017",
    year = "2016"
}

@article{Oliveira:2025kfg,
    author = "Oliveira, Vin{\'\i}cius and Escalona, Patricio and Angel, Lucia and de S. Pires, C. A. and Queiroz, Farinaldo S.",
    title = "{Type-II seesaw mechanism for Dirac neutrinos and its implications on N$_{eff}$ and lepton flavor violation in a 3-3-1 model}",
    eprint = "2502.01760",
    archivePrefix = "arXiv",
    primaryClass = "hep-ph",
    doi = "10.1007/JHEP07(2025)197",
    journal = "JHEP",
    volume = "07",
    pages = "197",
    year = "2025"
}

@article{Aguilar:2025grh,
    author = "Aguilar, Maria and Helo, Juan Carlos and Ota, Toshihiko and Queiroz, Farinaldo S. and Suarez, David and Rodr{\'\i}guez, Amanda",
    title = "{Type I + II Seesaw Model in light of the New Neutrino Oscillation Measurements}",
    eprint = "2509.23508",
    archivePrefix = "arXiv",
    primaryClass = "hep-ph",
    month = "9",
    year = "2025"
}

@article{DiFranzo:2013vra,
    author = "DiFranzo, Anthony and Nagao, Keiko I. and Rajaraman, Arvind and Tait, Tim M. P.",
    title = "{Simplified Models for Dark Matter Interacting with Quarks}",
    eprint = "1308.2679",
    archivePrefix = "arXiv",
    primaryClass = "hep-ph",
    reportNumber = "UCI-HEP-TR-2013-17, KEK-TH-1659",
    doi = "10.1007/JHEP11(2013)014",
    journal = "JHEP",
    volume = "11",
    pages = "014",
    year = "2013",
    note = "[Erratum: JHEP 01, 162 (2014)]"
}

@article{Berlin:2014tja,
    author = "Berlin, Asher and Hooper, Dan and McDermott, Samuel D.",
    title = "{Simplified Dark Matter Models for the Galactic Center Gamma-Ray Excess}",
    eprint = "1404.0022",
    archivePrefix = "arXiv",
    primaryClass = "hep-ph",
    reportNumber = "FERMILAB-PUB-14-060-A, MCTP-14-07",
    doi = "10.1103/PhysRevD.89.115022",
    journal = "Phys. Rev. D",
    volume = "89",
    number = "11",
    pages = "115022",
    year = "2014"
}

@article{Bell:2016uhg,
    author = "Bell, Nicole F. and Cai, Yi and Leane, Rebecca K.",
    title = "{Impact of mass generation for spin-1 mediator simplified models}",
    eprint = "1610.03063",
    archivePrefix = "arXiv",
    primaryClass = "hep-ph",
    doi = "10.1088/1475-7516/2017/01/039",
    journal = "JCAP",
    volume = "01",
    pages = "039",
    year = "2017"
}

@article{ElHedri:2017nny,
    author = "El Hedri, Sonia and Kaminska, Anna and de Vries, Maikel and Zurita, Jose",
    title = "{Simplified Phenomenology for Colored Dark Sectors}",
    eprint = "1703.00452",
    archivePrefix = "arXiv",
    primaryClass = "hep-ph",
    reportNumber = "MITP-17-002, TTP17-006",
    doi = "10.1007/JHEP04(2017)118",
    journal = "JHEP",
    volume = "04",
    pages = "118",
    year = "2017"
}

@article{Jacques:2015zha,
    author = {Jacques, Thomas and Nordstr{\"o}m, Karl},
    title = "{Mapping monojet constraints onto Simplified Dark Matter Models}",
    eprint = "1502.05721",
    archivePrefix = "arXiv",
    primaryClass = "hep-ph",
    doi = "10.1007/JHEP06(2015)142",
    journal = "JHEP",
    volume = "06",
    pages = "142",
    year = "2015"
}

@article{Backovic:2015soa,
    author = {Backovi{\'c}, Mihailo and Kr{\"a}mer, Michael and Maltoni, Fabio and Martini, Antony and Mawatari, Kentarou and Pellen, Mathieu},
    title = "{Higher-order QCD predictions for dark matter production at the LHC in simplified models with s-channel mediators}",
    eprint = "1508.05327",
    archivePrefix = "arXiv",
    primaryClass = "hep-ph",
    reportNumber = "MCNET-15-24, CP3-15-25, TTK-15-19",
    doi = "10.1140/epjc/s10052-015-3700-6",
    journal = "Eur. Phys. J. C",
    volume = "75",
    number = "10",
    pages = "482",
    year = "2015"
}

@article{Brennan:2016xjh,
    author = "Brennan, A. J. and McDonald, M. F. and Gramling, J. and Jacques, T. D.",
    title = "{Collide and Conquer: Constraints on Simplified Dark Matter Models using Mono-X Collider Searches}",
    eprint = "1603.01366",
    archivePrefix = "arXiv",
    primaryClass = "hep-ph",
    doi = "10.1007/JHEP05(2016)112",
    journal = "JHEP",
    volume = "05",
    pages = "112",
    year = "2016"
}

@article{BAEK201628,
title = {Beyond the dark matter effective field theory and a simplified model approach at colliders},
journal = {Physics Letters B},
volume = {756},
pages = {289-294},
year = {2016},
issn = {0370-2693},
doi = {https://doi.org/10.1016/j.physletb.2016.03.026},
url = {https://www.sciencedirect.com/science/article/pii/S0370269316001970},
author = {Seungwon Baek and P. Ko and Myeonghun Park and Wan-Il Park and Chaehyun Yu}
}

@article{PhysRevD.91.095020,
  title = {Searches for dark matter signals in simplified models at future hadron colliders},
  author = {Xiang, Qian-Fei and Bi, Xiao-Jun and Yin, Peng-Fei and Yu, Zhao-Huan},
  journal = {Phys. Rev. D},
  volume = {91},
  issue = {9},
  pages = {095020},
  numpages = {13},
  year = {2015},
  month = {May},
  publisher = {American Physical Society},
  doi = {10.1103/PhysRevD.91.095020},
  url = {https://link.aps.org/doi/10.1103/PhysRevD.91.095020}
}

@article{Bell_2016,
doi = {10.1088/1475-7516/2016/01/051},
url = {https://dx.doi.org/10.1088/1475-7516/2016/01/051},
year = {2016},
month = {jan},
publisher = {},
volume = {2016},
number = {01},
pages = {051},
author = {Nicole F. Bell and Yi Cai and Rebecca K. Leane},
title = {Mono-W dark matter signals at the LHC: simplified model analysis},
journal = {Journal of Cosmology and Astroparticle Physics}
}

@article{Longas:2023bvq,
    author = "Longas, Robinson and Rivera, Andres and Suarez, David and Ruiz, Cristian",
    title = "{Singlet{\textendash}doublet Dirac fermion dark matter from Peccei{\textendash}Quinn symmetry}",
    eprint = "2309.15052",
    archivePrefix = "arXiv",
    primaryClass = "hep-ph",
    doi = "10.1142/S0217751X24500994",
    journal = "Int. J. Mod. Phys. A",
    volume = "39",
    number = "25",
    pages = "2450099",
    year = "2024"
}

@article{Restrepo:2021kpq,
    author = "Restrepo, Diego and Suarez, David",
    title = "{Effective Dirac Neutrino Mass Operator in the Standard Model With a Local Abelian Extension}",
    eprint = "2112.09524",
    archivePrefix = "arXiv",
    primaryClass = "hep-ph",
    doi = "10.3389/fphy.2022.838531",
    journal = "Front. in Phys.",
    volume = "10",
    pages = "838531",
    year = "2022"
}

@article{Agudelo:2024luc,
    author = "Agudelo, Kimy and Restrepo, Diego and Rivera, Andr{\'e}s and Suarez, David",
    title = "{Multicomponent secluded WIMP dark matter and Dirac neutrino masses with an extra Abelian gauge symmetry}",
    eprint = "2412.02027",
    archivePrefix = "arXiv",
    primaryClass = "hep-ph",
    doi = "10.1103/PhysRevD.111.095018",
    journal = "Phys. Rev. D",
    volume = "111",
    number = "9",
    pages = "095018",
    year = "2025"
}

@article{Belanger:2001fz,
      author         = "Belanger, G. and Boudjema, F. and Pukhov, A. and Semenov,
                        A.",
      title          = "{MicrOMEGAs: A Program for calculating the relic density
                        in the MSSM}",
      journal        = "Comput. Phys. Commun.",
      volume         = "149",
      year           = "2002",
      pages          = "103-120",
      doi            = "10.1016/S0010-4655(02)00596-9",
      eprint         = "hep-ph/0112278",
      archivePrefix  = "arXiv",
      primaryClass   = "hep-ph",
      reportNumber   = "LAPTH-881-01",
      SLACcitation   = "%%CITATION = HEP-PH/0112278;%%"
}

@article{Belanger:2006is,
      author         = "Belanger, G. and Boudjema, F. and Pukhov, A. and Semenov,
                        A.",
      title          = "{MicrOMEGAs 2.0: A Program to calculate the relic density
                        of dark matter in a generic model}",
      journal        = "Comput. Phys. Commun.",
      volume         = "176",
      year           = "2007",
      pages          = "367-382",
      doi            = "10.1016/j.cpc.2006.11.008",
      eprint         = "hep-ph/0607059",
      archivePrefix  = "arXiv",
      primaryClass   = "hep-ph",
      reportNumber   = "LAPTH-1152-06",
      SLACcitation   = "%%CITATION = HEP-PH/0607059;%%"
}

@article{Belanger:2010gh,
      author         = "Belanger, G. and Boudjema, F. and Brun, P. and Pukhov, A.
                        and Rosier-Lees, S. and Salati, P. and Semenov, A.",
      title          = "{Indirect search for dark matter with micrOMEGAs2.4}",
      journal        = "Comput. Phys. Commun.",
      volume         = "182",
      year           = "2011",
      pages          = "842-856",
      doi            = "10.1016/j.cpc.2010.11.033",
      eprint         = "1004.1092",
      archivePrefix  = "arXiv",
      primaryClass   = "hep-ph",
      reportNumber   = "IRFU-10-24, LAPTH-012-10.",
      SLACcitation   = "%%CITATION = ARXIV:1004.1092;%%"
}

@article{Belanger:2013oya,
      author         = "Belanger, G. and Boudjema, F. and Pukhov, A. and Semenov,
                        A.",
      title          = "{micrOMEGAs$\_3$: A program for calculating dark matter
                        observables}",
      journal        = "Comput. Phys. Commun.",
      volume         = "185",
      year           = "2014",
      pages          = "960-985",
      doi            = "10.1016/j.cpc.2013.10.016",
      eprint         = "1305.0237",
      archivePrefix  = "arXiv",
      primaryClass   = "hep-ph",
      reportNumber   = "LAPTH-023-13",
      SLACcitation   = "%%CITATION = ARXIV:1305.0237;%%"
}

@article{Belanger:2014vza,
      author         = "B\'elanger, G. and Boudjema, F. and Pukhov, A. and
                        Semenov, A.",
      title          = "{micrOMEGAs4.1: two dark matter candidates}",
      journal        = "Comput. Phys. Commun.",
      volume         = "192",
      year           = "2015",
      pages          = "322-329",
      doi            = "10.1016/j.cpc.2015.03.003",
      eprint         = "1407.6129",
      archivePrefix  = "arXiv",
      primaryClass   = "hep-ph",
      SLACcitation   = "%%CITATION = ARXIV:1407.6129;%%"
}

@article{Belyaev:2012qa,
      author         = "Belyaev, Alexander and Christensen, Neil D. and Pukhov,
                        Alexander",
      title          = "{CalcHEP 3.4 for collider physics within and beyond the
                        Standard Model}",
      journal        = "Comput. Phys. Commun.",
      volume         = "184",
      year           = "2013",
      pages          = "1729-1769",
      doi            = "10.1016/j.cpc.2013.01.014",
      eprint         = "1207.6082",
      archivePrefix  = "arXiv",
      primaryClass   = "hep-ph",
      reportNumber   = "PITT-PACC-1209",
      SLACcitation   = "%%CITATION = ARXIV:1207.6082;%%"
}

@article{Alguero:2023zol,
    author = "Alguero, G. and Belanger, G. and Boudjema, F. and Chakraborti, S. and Goudelis, A. and Kraml, S. and Mjallal, A. and Pukhov, A.",
    title = "{micrOMEGAs 6.0: N-component dark matter}",
    eprint = "2312.14894",
    archivePrefix = "arXiv",
    primaryClass = "hep-ph",
    doi = "10.1016/j.cpc.2024.109133",
    journal = "Comput. Phys. Commun.",
    volume = "299",
    pages = "109133",
    year = "2024"
}

@book{CTAConsortium:2017dvg,
    author = "Acharya, B. S. and others",
    collaboration = "CTA Consortium",
    title = "{Science with the Cherenkov Telescope Array}",
    eprint = "1709.07997",
    archivePrefix = "arXiv",
    primaryClass = "astro-ph.IM",
    doi = "10.1142/10986",
    isbn = "978-981-327-008-4",
    publisher = "WSP",
    month = "11",
    year = "2018"
}

@article{Profumo:2013sca,
    author = "Profumo, Stefano and Queiroz, Farinaldo S.",
    title = "{Constraining the $Z'$ mass in 331 models using direct dark matter detection}",
    eprint = "1307.7802",
    archivePrefix = "arXiv",
    primaryClass = "hep-ph",
    reportNumber = "CETUP2013-012",
    doi = "10.1140/epjc/s10052-014-2960-x",
    journal = "Eur. Phys. J. C",
    volume = "74",
    number = "7",
    pages = "2960",
    year = "2014"
}

@article{Arcadi:2017wqi,
    author = "Arcadi, Giorgio and Lindner, Manfred and Queiroz, Farinaldo S. and Rodejohann, Werner and Vogl, Stefan",
    title = "{Pseudoscalar Mediators: A WIMP model at the Neutrino Floor}",
    eprint = "1711.02110",
    archivePrefix = "arXiv",
    primaryClass = "hep-ph",
    doi = "10.1088/1475-7516/2018/03/042",
    journal = "JCAP",
    volume = "03",
    pages = "042",
    year = "2018"
}

@article{Gelmini:2006mr,
    author = "Gelmini, Graciela B. and Gondolo, Paolo and Soldatenko, Adrian and Yaguna, C. E.",
    title = "{Direct detection of neutralino dark mattter in non-standard cosmologies}",
    eprint = "hep-ph/0610379",
    archivePrefix = "arXiv",
    doi = "10.1103/PhysRevD.76.015010",
    journal = "Phys. Rev. D",
    volume = "76",
    pages = "015010",
    year = "2007"
}

@article{Gelmini:2006pw,
    author = "Gelmini, Graciela B. and Gondolo, Paolo",
    title = "{Neutralino with the right cold dark matter abundance in (almost) any supersymmetric model}",
    eprint = "hep-ph/0602230",
    archivePrefix = "arXiv",
    reportNumber = "UCLA-06-TEP-07",
    doi = "10.1103/PhysRevD.74.023510",
    journal = "Phys. Rev. D",
    volume = "74",
    pages = "023510",
    year = "2006"
}

@article{Arcadi:2011ev,
    author = "Arcadi, Giorgio and Ullio, Piero",
    title = "{Accurate estimate of the relic density and the kinetic decoupling in non-thermal dark matter models}",
    eprint = "1104.3591",
    archivePrefix = "arXiv",
    primaryClass = "hep-ph",
    doi = "10.1103/PhysRevD.84.043520",
    journal = "Phys. Rev. D",
    volume = "84",
    pages = "043520",
    year = "2011"
}

@article{Baer:2014eja,
    author = "Baer, Howard and Choi, Ki-Young and Kim, Jihn E. and Roszkowski, Leszek",
    title = "{Dark matter production in the early Universe: beyond the thermal WIMP paradigm}",
    eprint = "1407.0017",
    archivePrefix = "arXiv",
    primaryClass = "hep-ph",
    doi = "10.1016/j.physrep.2014.10.002",
    journal = "Phys. Rept.",
    volume = "555",
    pages = "1--60",
    year = "2015"
}

@article{Drukier:1986tm,
    author = "Drukier, A. K. and Freese, Katherine and Spergel, D. N.",
    title = "{Detecting Cold Dark Matter Candidates}",
    doi = "10.1103/PhysRevD.33.3495",
    journal = "Phys. Rev. D",
    volume = "33",
    pages = "3495--3508",
    year = "1986"
}

@article{Fan:2010gt,
    author = "Fan, JiJi and Reece, Matthew and Wang, Lian-Tao",
    title = "{Non-relativistic effective theory of dark matter direct detection}",
    eprint = "1008.1591",
    archivePrefix = "arXiv",
    primaryClass = "hep-ph",
    doi = "10.1088/1475-7516/2010/11/042",
    journal = "JCAP",
    volume = "11",
    pages = "042",
    year = "2010"
}

@article{Chun:2016cnm,
    author = "Chun, Eung Jin and Jung, Sunghoon and Park, Jong-Chul",
    title = "{Very Degenerate Higgsino Dark Matter}",
    eprint = "1607.04288",
    archivePrefix = "arXiv",
    primaryClass = "hep-ph",
    doi = "10.1007/JHEP01(2017)009",
    journal = "JHEP",
    volume = "01",
    pages = "009",
    year = "2017"
}

@article{Griest:2000kj,
    author = "Griest, K. and Kamionkowski, M.",
    title = "{Supersymmetric dark matter}",
    doi = "10.1016/S0370-1573(00)00021-1",
    journal = "Phys. Rept.",
    volume = "333",
    pages = "167--182",
    year = "2000"
}

@article{Arbelaez:2015ila,
    author = "Arbelaez, Carolina and Longas, Robinson and Restrepo, Diego and Zapata, Oscar",
    title = "{Fermion dark matter from SO(10) GUTs}",
    eprint = "1509.06313",
    archivePrefix = "arXiv",
    primaryClass = "hep-ph",
    doi = "10.1103/PhysRevD.93.013012",
    journal = "Phys. Rev. D",
    volume = "93",
    number = "1",
    pages = "013012",
    year = "2016"
}

@article{Ellis:1988bf,
    author = "Ellis, John R. and Hagelin, John S. and Kelley, Stephen and Nanopoulos, Dimitri V. and Olive, Keith A.",
    title = "{FLIPPED DARK MATTER}",
    reportNumber = "MIU-THP-88/026, MAD/TH/88-7, CERN-TH-4989/88, LBL-24922, UMN-TH-649/88",
    doi = "10.1016/0370-2693(88)90947-1",
    journal = "Phys. Lett. B",
    volume = "209",
    pages = "283--288",
    year = "1988"
}

@article{Kadastik:2009cu,
    author = "Kadastik, Mario and Kannike, Kristjan and Raidal, Martti",
    title = "{Dark Matter as the signal of Grand Unification}",
    eprint = "0907.1894",
    archivePrefix = "arXiv",
    primaryClass = "hep-ph",
    doi = "10.1103/PhysRevD.80.085020",
    journal = "Phys. Rev. D",
    volume = "80",
    pages = "085020",
    year = "2009",
    note = "[Erratum: Phys.Rev.D 81, 029903 (2010)]"
}

@article{Lindner:2016bgg,
    author = "Lindner, Manfred and Platscher, Moritz and Queiroz, Farinaldo S.",
    title = "{A Call for New Physics : The Muon Anomalous Magnetic Moment and Lepton Flavor Violation}",
    eprint = "1610.06587",
    archivePrefix = "arXiv",
    primaryClass = "hep-ph",
    doi = "10.1016/j.physrep.2017.12.001",
    journal = "Phys. Rept.",
    volume = "731",
    pages = "1--82",
    year = "2018"
}

@article{Toma:2013zsa,
    author = "Toma, Takashi and Vicente, Avelino",
    title = "{Lepton Flavor Violation in the Scotogenic Model}",
    eprint = "1312.2840",
    archivePrefix = "arXiv",
    primaryClass = "hep-ph",
    reportNumber = "DCPT-13-198",
    doi = "10.1007/JHEP01(2014)160",
    journal = "JHEP",
    volume = "01",
    pages = "160",
    year = "2014"
}

@article{Lychkovskiy:2010ue,
    author = "Lychkovskiy, Oleg and Blinnikov, Sergei and Vysotsky, Mikhail",
    title = "{TeV-scale bileptons, see-saw type II and lepton flavor violation in core-collapse supernova}",
    eprint = "0912.1395",
    archivePrefix = "arXiv",
    primaryClass = "hep-ph",
    doi = "10.1140/epjc/s10052-010-1291-9",
    journal = "Eur. Phys. J. C",
    volume = "67",
    pages = "213--227",
    year = "2010"
}

@article{Dinh:2012bp,
    author = "Dinh, D. N. and Ibarra, A. and Molinaro, E. and Petcov, S. T.",
    title = "{The $\mu - e$ Conversion in Nuclei, $\mu \to e \gamma, \mu \to 3e$ Decays and TeV Scale See-Saw Scenarios of Neutrino Mass Generation}",
    eprint = "1205.4671",
    archivePrefix = "arXiv",
    primaryClass = "hep-ph",
    reportNumber = "FLAVOUR(267104)-ERC-15, SISSA-10-2012-EP, TUM-HEP-837-12, CFTP-12-007",
    doi = "10.1007/JHEP08(2012)125",
    journal = "JHEP",
    volume = "08",
    pages = "125",
    year = "2012",
    note = "[Erratum: JHEP 09, 023 (2013)]"
}

@article{He:2014efa,
    author = "He, Xiao-Gang and Lee, Chao-Jung and Tandean, Jusak and Zheng, Ya-Juan",
    title = "{Seesaw Models with Minimal Flavor Violation}",
    eprint = "1411.6612",
    archivePrefix = "arXiv",
    primaryClass = "hep-ph",
    doi = "10.1103/PhysRevD.91.076008",
    journal = "Phys. Rev. D",
    volume = "91",
    number = "7",
    pages = "076008",
    year = "2015"
}

@article{Lavoura:2003xp,
    author = "Lavoura, L.",
    title = "{General formulae for f(1) ---{\ensuremath{>}} f(2) gamma}",
    eprint = "hep-ph/0302221",
    archivePrefix = "arXiv",
    doi = "10.1140/epjc/s2003-01212-7",
    journal = "Eur. Phys. J. C",
    volume = "29",
    pages = "191--195",
    year = "2003"
}

@article{Miscetti:2025uxk,
    author = "Miscetti, S.",
    collaboration = "Mu2e",
    title = "{Status of the Mu2e experiment}",
    reportNumber = "FERMILAB-PUB-25-0075-T",
    doi = "10.1016/j.nima.2025.170257",
    journal = "Nucl. Instrum. Meth. A",
    volume = "1073",
    pages = "170257",
    year = "2025"
}

@article{Arganda:2007jw,
    author = "Arganda, E. and Herrero, M. J. and Teixeira, A. M.",
    title = "{mu-e conversion in nuclei within the CMSSM seesaw: Universality versus non-universality}",
    eprint = "0707.2955",
    archivePrefix = "arXiv",
    primaryClass = "hep-ph",
    reportNumber = "FTUAM-07-10, IFT-UAM-CSIC-07-28, LPT-ORSAY-07-51",
    doi = "10.1088/1126-6708/2007/10/104",
    journal = "JHEP",
    volume = "10",
    pages = "104",
    year = "2007"
}

@article{SINDRUMII:2006dvw,
    author = "Bertl, Wilhelm H. and others",
    collaboration = "SINDRUM II",
    title = "{A Search for muon to electron conversion in muonic gold}",
    doi = "10.1140/epjc/s2006-02582-x",
    journal = "Eur. Phys. J. C",
    volume = "47",
    pages = "337--346",
    year = "2006"
}

@article{Ferreira:2011hm,
    author = "Ferreira, Jr, J. G. and Pinheiro, P. R. D. and Pires, C. A. de S. and da Silva, P. S. Rodrigues",
    title = "{The Minimal 3-3-1 model with only two Higgs triplets}",
    eprint = "1109.0031",
    archivePrefix = "arXiv",
    primaryClass = "hep-ph",
    doi = "10.1103/PhysRevD.84.095019",
    journal = "Phys. Rev. D",
    volume = "84",
    pages = "095019",
    year = "2011"
}

@article{Phong:2013cfa,
    author = "Phong, Vo Quoc and Van, Vo Thanh and Long, Hoang Ngoc",
    title = "{Electroweak phase transition in the reduced minimal 3-3-1 model}",
    eprint = "1309.0355",
    archivePrefix = "arXiv",
    primaryClass = "hep-ph",
    doi = "10.1103/PhysRevD.88.096009",
    journal = "Phys. Rev. D",
    volume = "88",
    pages = "096009",
    year = "2013"
}

@article{Escalona:2025rxu,
    author = "Escalona, Patricio and Pinheiro, Jo{\~a}o Paulo and Doff, A. and de S. Pires, C. A.",
    title = "{Meson mixing bounds on Z$^{\prime}$ mass in the alignment limit: establishing the phenomenological viability of the 331 model}",
    eprint = "2503.14653",
    archivePrefix = "arXiv",
    primaryClass = "hep-ph",
    doi = "10.1007/JHEP07(2025)105",
    journal = "JHEP",
    volume = "07",
    pages = "105",
    year = "2025"
}

@article{Escalona:2025jla,
    author = "Escalona, Patricio and Pinheiro, Jo{\~a}o Paulo and Oliveira, Vin{\'\i}cius and Doff, Adriano and De Sousa Pires, Carlos Antonio",
    title = "{Three Decades of FCNC Studies in 3-3-1 Model with Right-Handed Neutrinos: From Z'-Dominance to the Alignment Limit}",
    eprint = "2510.17979",
    archivePrefix = "arXiv",
    primaryClass = "hep-ph",
    doi = "10.3390/universe11120396",
    journal = "Universe",
    volume = "11",
    number = "12",
    pages = "396",
    year = "2025"
}

@article{Doff:2026kcs,
    author = "Doff, A. and Pires, C. A. de S.",
    title = "{Scalar contributions to the S, T, U parameters in a 3-3-1 model}",
    eprint = "2603.08980",
    archivePrefix = "arXiv",
    primaryClass = "hep-ph",
    month = "3",
    year = "2026"
}

@article{Kannike:2025qru,
    author = "Kannike, Kristjan and Koivunen, Niko and Kubarski, Aleksei",
    title = "{Is our vacuum global in a 331 model with three triplets?}",
    eprint = "2509.18250",
    archivePrefix = "arXiv",
    primaryClass = "hep-ph",
    doi = "10.1007/JHEP01(2026)115",
    journal = "JHEP",
    volume = "01",
    pages = "115",
    year = "2026"
}

@article{Nevzorov:2025ido,
    author = "Nevzorov, Roman",
    title = "{Smallness of neutrino masses and leptogenesis in 331 composite Higgs model}",
    eprint = "2509.13245",
    archivePrefix = "arXiv",
    primaryClass = "hep-ph",
    doi = "10.1016/j.physletb.2026.140210",
    journal = "Phys. Lett. B",
    volume = "874",
    pages = "140210",
    year = "2026"
}

@article{Rehman:2025urc,
    author = "Rehman, Muhammad and Iqbal, Muhammad Adeel and Gomez, Mario E. and Panella, Orlando",
    title = "{Radiative corrections to the S, T, U parameters and their impact on the W boson mass in the 331 model}",
    eprint = "2507.18527",
    archivePrefix = "arXiv",
    primaryClass = "hep-ph",
    doi = "10.1103/vx7q-ns9p",
    journal = "Phys. Rev. D",
    volume = "112",
    number = "5",
    pages = "055029",
    year = "2025"
}

@article{Schumann:2015cpa,
    author = {Schumann, Marc and Baudis, Laura and B{\"u}tikofer, Lukas and Kish, Alexander and Selvi, Marco},
    title = "{Dark matter sensitivity of multi-ton liquid xenon detectors}",
    eprint = "1506.08309",
    archivePrefix = "arXiv",
    primaryClass = "physics.ins-det",
    doi = "10.1088/1475-7516/2015/10/016",
    journal = "JCAP",
    volume = "10",
    pages = "016",
    year = "2015"
}

@article{Billard:2021uyg,
    author = "Billard, Julien and others",
    title = "{Direct detection of dark matter{\textemdash}APPEC committee report*}",
    eprint = "2104.07634",
    archivePrefix = "arXiv",
    primaryClass = "hep-ex",
    doi = "10.1088/1361-6633/ac5754",
    journal = "Rept. Prog. Phys.",
    volume = "85",
    number = "5",
    pages = "056201",
    year = "2022"
}
